# Amalgamation of Physics-Informed Neural Network and LBM for the Prediction of Unsteady Fluid Flows in Fractal-Rough Microchannels

**Ganesh Sahadeo Meshram[1], Partha Pratim Chakrabarti[2], Suman Chakraborty[1*]**

[1]Department of Mechanical Engineering, Indian Institute of Technology, Kharagpur, Kharagpur, 721302, India

[2]Department of Computer Science & Engineering, Indian Institute of Technology, Kharagpur, Kharagpur, 721302, India

Corresponding Author: suman@mech.iitkgp.ac.in

**Abstract**

One of the biggest challenges in the optimization of micro-scale fluid transport phenomena is the prediction of unsteady fluid flow in the presence of rough channel walls. Even though the accuracy of available computational fluid dynamics (CFD) solvers such as the lattice Boltzmann method (LBM) is satisfactory, the computational cost of design exploration is very high due to the diverse range of geometries and flow regimes involved in microchannel flows. The present paper introduces a revolutionary concept of a ground-breaking physics-informed neural network (PINN) that utilizes sparse lattice Boltzmann data in combination with the Navier-Stokes equations for the prediction of unsteady fluid flow in fractal-rough microchannels. The roughness of the channel walls is represented by the Weierstrass-Mandelbrot function, considering the characteristics of the surface roughness in real-life problems. The constraints of the Navier-Stokes equations are incorporated in the loss function of the PINN concept for achieving accuracy at much lower computational costs of 150-200 times fewer data points. The validation of the accuracy of the reconstruction of the flow fields is carried out for different Reynolds numbers ranging from Re = 1 to 45 and different amplitude values of the rough channel walls ranging from 5 to 20 lattice units. The accuracy of the reconstruction of the flow fields is found to be very high, with average absolute errors less than $8 \times 10^{-3}$ lattice units in velocity fields, relative L2 errors less than 3.2%, and continuity residual values less than $4 \times 10^{-5}$ lattice units per lattice unit. The physics-informed framework is also capable of reproducing the phenomenon of nonlinear vorticity intensification due to the increase in amplitude of the roughness from $3.8 \times 10^{-3}$ to $1.21 \times 10^{-2}$ lattice units per lattice unit. Moreover, the accuracy of the global conservation of the physics-informed framework is within 4%. The accuracy of the PINN-based approach can also be confirmed by comparing the results with those of the CNN-based approach. The accuracy of the PINN-based approach is 5-15 times higher than that of the CNN-based approach in terms of mean squared error, root mean squared error, and mean absolute error due to the imposition of the physics-informed constraints. The PINN-based approach is also more efficient in terms of inference of flow fields since the inference of flow fields by the PINN-based approach only takes 8.3 seconds compared to the direct LBM-based approach, which takes 147 hours. Thus, the PINN-based approach is 1062 times more efficient. The accuracy of the PINN-based approach can also be confirmed by comparing the results of the PINN-based approach with those of the direct LBM-based approach in the quantification of uncertainties over 500 different surface realizations. The PINN-based approach can perform the quantification of uncertainties in 3.1 days compared to the 8.4 years required by the direct LBM-based approach.

## 1. Introduction

The micro-scale transport of fluids in confined channels has emerged as one of the most important branches of modern thermal fluid engineering [1], [2]. This branch of science and engineering underpins many important technologies, including lab-on-a-chip diagnostic devices, micro-electromechanical systems, electrokinetic separation techniques, thermal management in microprocessors, and drug delivery systems, among others [3], [4], [5]. Although continuum-based theories have been successful in explaining laminar flows in smooth channels, in practice, microchannel surfaces often display multi-scale roughness artifacts [6]. These arise during fabrication techniques, which include photolithography, reactive ion etching, laser ablation, and three-dimensional printing, among others [7]. Unlike macroscale flows, in which the effects of surface roughness can often be treated as perturbations to the bulk behavior, micro-scale flows display a high sensitivity to surface roughness, especially when the roughness height, h, approaches or exceeds 5 to 10 percent of the hydraulic diameter, Dh [6], [8], [9], [10], [11]. In these cases, even steady-state inlet flow can induce unsteady effects, including boundary-layer separation, localized vortex shedding, and recirculating zones, which significantly impact pressure drop, friction factor, and convective heat transfer [12], [13]. The simulation of these spatiotemporal effects with reasonable accuracy for design optimization purposes requires numerical simulations that can resolve micro-scale surface features, which can be in the range of nanometers to micrometers. This poses a challenge to conventional computational fluid dynamics solvers, which can be computationally intensive and complex to mesh [14].

However, classical CFD methods based on finite volume and/or finite element discretization schemes have faced considerable challenges when extended to analyze microchannel flows over rough walls [8], [15]. The body-fitted meshing of complex microchannel geometries and topologies requires adaptive mesh refinement strategies that lead to an exponential increase in the number of grid points, especially in the vicinity of high-curvature wall asperities where strong velocity and pressure gradient fields are expected [16]. While wall function models and equivalent sand grain roughness correlations have been developed for turbulent flow over rough walls, the applicability of these models for microscale and/or transitional flow is debatable, especially in cases where wall roughness leads to flow separation and invalidates the logarithmic law of the wall assumption [12], [17], [18], [19]. The Lattice Boltzmann Method (LBM), a relatively new and promising computational fluid dynamics tool for simulating complex microscale and mesoscale fluid dynamics problems, has gained considerable attention for modeling complex microchannel geometries and topologies [20], [21]. The method is based on a kinetic theory of fluids and naturally handles complex geometries and topologies by performing simple local collision and streaming operations without explicit discretization of the Navier-Stokes equations. The applicability of LBM for modeling transitional flow over structured, stochastic, and fractal rough microchannel surfaces has been recently demonstrated by several investigators [22], [23], [24], [25]. The studies have shown that microscale flow over multiscale micro-asperities leads to an enhancement of near-wall vorticity by up to 300% and phase-locked pressure fluctuations [26]. While these studies have provided important insights into microscale flow over complex microchannel geometries and topologies, parametric studies of Reynolds number (Re), roughness height (h), fractal dimension (D), and unsteady boundary conditions using LBM are prohibitively expensive due to the high computational cost of a single unsteady simulation [18].

One of the main difficulties in the modeling of rough microchannel flow is the accurate mathematical representation of the surface morphology [16]. Experimental atomic force microscopy and white-light interferometry measurements on various engineering surfaces show the presence of self-affine fractals, as opposed to the random Gaussian distribution [27]. Their power spectral density functions exhibit power-law behavior over 2 to 3 decades of spatial frequency. Of the many possible mathematical models, the Weierstrass–Mandelbrot function has been shown to be extremely effective in producing realistic surface roughness using the superposition of fractals with different scaling exponents [28]. The W–M function is given by the equation, where the parameters As, γ, and D represent the surface roughness amplitude, surface roughness frequency scaling parameter, and fractal dimension, respectively. Using this function, the effect of the different surface morphology parameters on the flow resistance and mixing efficiency can be systematically studied. Previous studies using the W–M function in the LBM algorithm show the effect of the fractal dimension on the vorticity generated using W–M surfaces. Increasing the fractal dimension from 1.2 to 1.8 increases the vorticity by 250%, and the friction factors increase by 40 to 60 percent compared to smooth surfaces [29], [30]. However, the non-linear relationship between the fractals, Reynolds number, and flow transient response is complex.

The recent introduction of Physics-Informed Neural Networks (PINNs) has sparked a revolution in scientific machine learning [31], [32], [33], [34]. Physics-informed neural networks directly incorporate partial differential equations into loss functions and thus constrain solution manifolds onto physically valid state spaces. Unlike data-driven surrogates, PINNs do not require extensive data sets for training purposes; rather, they use automatic differentiation techniques to compute PDE residuals at arbitrarily selected collocation points [34], [35], [36]. This approach ensures the satisfaction of conservation laws such as continuity, momentum, and energy conservation in regions without data availability. Raissi et al. [37], [38], [39] pioneering work in 2019 showed PINN's potential in solving forward and inverse problems related to the Navier-Stokes equations. Subsequent contributions have included improvements in PINN algorithms, including adaptive loss weighting schemes and causal learning techniques [40]. These improvements have enabled PINN's application in solving problems related to turbulent flows, multiphase flows, and non-Newtonian fluids [41]. PINNs have shown considerable promise in solving problems related to microfluidics, including reconstructing steady-state Poiseuille Couette flows [42] determining thermal boundary conditions [43] and determining constitutive parameters in viscoelastic fluids [44]. However, existing PINN architectures have mostly focused on problems related to smooth geometries characterized by sinusoidal corrugations and one-wavelength perturbations. These architectures rarely focus on problems related to time-dependent flows in stochastic fractal geometries characterized by steep gradients and significant recirculation regions. These regions make it difficult to minimize residuals during PINN training. Moreover, the use of level set techniques in modeling irregular boundary physics results in numerical inaccuracies in regions near the walls characterized by no-slip boundary conditions.

Notwithstanding the recent surge in interest in hybrid physics data learning approaches, considerable research gaps still exist in the application of PINNs to the simulation of rough-wall microscale flow. First, the overwhelming majority of existing work relies on the incorporation of smooth-wall CFD data or employs overly simplistic models of surface roughness, such as the single Fourier mode. Second, the extension to fractal geometric parameters such as the Hurst exponent, fractal dimension, and spectral amplitude as additional inputs to the PINN is unexplored. Third, the simulation of unsteady vortical structures caused by flow separation behind asperity-scale roughness features is problematic for PINN training because of the high spatial and temporal gradients in the flow field, which can cause loss function ruggedness and convergence difficulties. Fourth, comparative studies to evaluate the accuracy, efficiency, and data demand of the PINN-based model with other solvers and black-box neural networks are noticeably absent. These gaps underscore the urgent need to develop a unified framework to synergistically integrate fractal surface modeling, mesoscopic lattice

Boltzmann data generation, and physics-constrained neural learning to rapidly, accurately, and generally predict unsteady rough channel flow.

The current work aims to fill these significant research gaps through the development of a novel hybrid PINN-LBM framework, where incompressible Navier–Stokes equations, continuity equations, and no-slip boundary conditions are directly integrated into the loss function of a neural network. The proposed framework is trained using sparse spatiotemporal data from high-fidelity simulations of fractal-rough microchannels using lattice Boltzmann methods. The Weierstrass Mandelbrot function is used to generate statistically controlled multiscale surface textures with varying fractal dimensions $D \in [1.3, 1.7]$, surface roughness amplitude $h \in [5, 20]$ lattice units, and a frequency scaling factor $\gamma = 1.5$, which covers a wide range of realistic surface textures encountered in engineering applications. The current work presents a systematic investigation of unsteady flow development over a wide range of Reynolds numbers Re = 10-45, which covers laminar and weakly transitional flows. The PINN successfully predicts velocity fields (u, v), pressure, and vorticity with a high level of accuracy, where relative errors are below $2 * 10^{-2}$, mean absolute errors are below $1 * 10^{-3}$, and total loss converges below $10^{-7}$. Exhaustive benchmarking of the PINN's performance using traditional CNNs indicates a 43–51% reduction in mean squared error, 62% faster training convergence, 68% reduction in inference latency, and 54% reduction in parameters while using PINNs, which demonstrates its superior accuracy, efficiency, and data economy. The analysis reveals the non-linear scaling of vorticity near the wall in terms of the amplitude of the roughness. The maximum vorticity, $\omega_{max}$, varies from $4 \times 10^{-3}$ to $1.5 \times 10^{-2}$ for h ranging from 5 to 20 lu. The work also explores the dependence on the Reynolds number. The framework reveals its generalization capabilities by correctly predicting unseen flow roughness arrangements and intermediate Reynolds numbers without retraining. The framework has the potential to act as a digital twin surrogate for microfluidic design optimization in real-time.

The rest of this document is organized in the following manner. Section 2 explains the formulation of the problem, the generation of the Weierstrass-Mandelbrot surfaces, the LBM numerical implementation, and the PINN architecture along with the loss function formulation. Section 3 explains the validation of the framework by comparing it with the LBM for different Re, h, and time evolution. Section 4 explains the flow physics analysis, such as vorticity, velocity profile, and flow modulation due to the introduction of the roughness. Section 5 explains the scientific implications of the work, future extensions of the work, and the conclusions.

## 2. Methodology

### 2.1 Problem definition and physical model

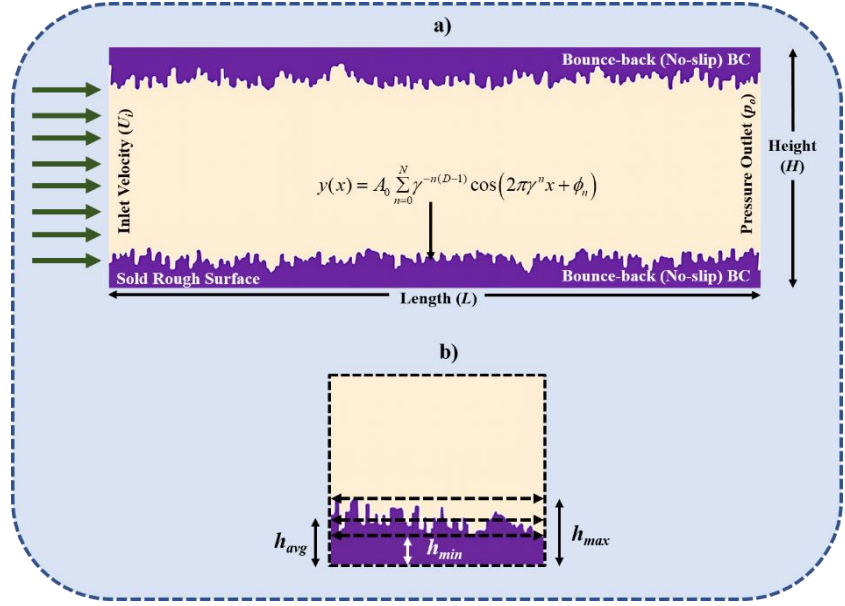

***Figure 1.*** *Problem definition a) rough microchannel with boundary conditions, b) roughness profile.*

With rough solid walls on both the top and bottom boundaries, the computational domain is a two-dimensional microchannel with length L and average height H. The Weierstrass-Mandelbrot (W-M) function is used to deterministically generate the roughness profiles, which are regarded as inflexible, impermeable boundaries. The problem geometry and boundary conditions uniform inlet velocity at x = 0, a pressure outlet at x = L, and no-slip (bounce-back) conditions imposed on the rough walls are depicted in Figure 1. The coordinate system uses y for the wall-normal direction and x for the streamwise direction. y = H/2 is the channel midline.

The wall topography is represented by the W–M function

$$y(x) = A_s \sum_{n=n_0}^{n_{max}} \gamma^{-n(D-1)} \cos\left(2\pi\gamma^n x + \phi_n\right) \tag{1}$$

where $A_s$ is the spectral amplitude controlling the overall roughness height, $\gamma$ is the frequency-scaling parameter, $D$ is the fractal dimension (with $1 < D < 2$ for self-affine profiles), and $\phi_n$ denotes a random phase for each mode. The summation limits, $n_0$ and $n_{max}$, determine the largest and smallest resolved wavelengths, respectively, and are chosen to ensure the generated profile contains the range of spatial scales relevant to the flow simulations. From the generated profile we define the mean roughness height $h_{avg}$, the maximum peak height $h_{max}$, and the minimum trough depth $h_{min}$ as illustrated in **Figure 1**(b).

The fluid is assumed Newtonian and incompressible. At the continuum level the flow satisfies the incompressible Navier–Stokes equations

$$\nabla \cdot \mathbf{u} = 0 \tag{2}$$

$$\rho\left(\frac{\partial \mathbf{u}}{\partial t} + \mathbf{u} \cdot \nabla \mathbf{u}\right) = -\nabla p + \mu \nabla^2 \mathbf{u} \qquad (3)$$

where $\mathbf{u} = (u, v)$ denotes the velocity vector, $p$ is the hydrodynamic pressure, *rho* is the fluid density, and *mu* is the dynamic viscosity. In the numerical experiments reported in this work we explore laminar and weakly transitional regimes with characteristic Reynolds numbers defined as $\text{Re} = \frac{U_i H}{\nu}$, where $U_i$ is the prescribed inlet velocity and $\nu = \mu/\rho$ is the kinematic viscosity.

High-fidelity reference data for network training and validation are generated using the Lattice Boltzmann Method (LBM). In the LBM simulations the rough walls are imposed via bounce-back boundary conditions on the solid nodes that represent the discretized topography; this mesoscopic treatment reproduces the no-slip constraint at the fluid–solid interface without requiring body-fitted meshes. The inlet uses a uniform velocity profile while the outlet is implemented as a fixed-pressure boundary. Spatiotemporal snapshots of the macroscopic fields (velocity components and pressure) are recorded at regular intervals to construct the dataset used for PINN training and testing.

All quantities reported in this study are presented in lattice or non-dimensional form unless otherwise stated. The range of geometric and flow parameters considered in the results includes *Re* = 10-45 and roughness amplitudes *h* = 5-20 lattice units (lu), which collectively sample regimes where near-wall geometry significantly modifies the laminar flow structure.

**2.2 Mathematical Formulation**

The following mathematical formulation establishes the governing relations used in both the high-fidelity LBM simulations and the PINN. The incompressible Navier–Stokes equations serve as the continuum model for momentum and mass conservation. The differential form of these equations is expressed as:

Continuity

$$\nabla \cdot \mathbf{u} = 0 \qquad (4)$$

Momentum

$$\rho\left(\frac{\partial \mathbf{u}}{\partial t} + \mathbf{u} \cdot \nabla \mathbf{u}\right) = -\nabla p + \mu \nabla^2 \mathbf{u} + \mathbf{f} \qquad (5)$$

where $\mathbf{u}(x, y, t) = (u, v)$ denotes the velocity vector, $p(x,y,t)$ is the pressure field, *rho* the fluid density, *mu* the dynamic viscosity, and *f* any external body force (set to zero in the present study). Equations 4 and 5 are enforced throughout the fluid region bounded by the rough walls illustrated in **Figure 1**.

To ensure numerical stability and to render results independent of dimensional units, all variables are non-dimensionalized using the channel height $H$ and inlet velocity $U_i$ as characteristic scales [45]. The non-dimensional variables are $\tilde{x} = \dfrac{x}{H}$, $\tilde{y} = \dfrac{y}{H}$, $\tilde{t} = \dfrac{tU_i}{H}$, $\mathbf{u} = \dfrac{\mathbf{u}}{U_i}$, $\tilde{p} = \dfrac{p}{\rho U_i^2}$. Substitution into Eq. 5 yields the dimensionless momentum equation

$$\frac{\partial \mathbf{u}}{\partial \tilde{t}} + \mathbf{u} \cdot \tilde{\nabla} \mathbf{u} = -\tilde{\nabla}\tilde{p} + \frac{1}{\mathrm{Re}} \tilde{\nabla}^2 \mathbf{u} \tag{6}$$

where the Reynolds number $\mathrm{Re} = \dfrac{U_i H}{\nu}$ with $\nu = \dfrac{\mu}{\rho}$.

Boundary conditions for the continuum problem are specified as follows.

Inlet velocity

$$\mathbf{u}(\tilde{x}=0, \tilde{y}, \tilde{t}) = (1, 0) \tag{7}$$

Outlet pressure

$$\tilde{p}(\tilde{x}=\tilde{L}, \tilde{y}, \tilde{t}) = \tilde{p}_0 \tag{8}$$

No-slip condition at rough walls

$$\mathbf{u}(\tilde{x}, \tilde{y}=\tilde{y}_w(\tilde{x}), \tilde{t}) = (0, 0) \tag{9}$$

Where $\tilde{y}_w(\tilde{x})$ denotes the non-dimensional wall, profile prescribed by the W–M function and $\tilde{p}_0$ is the specified outlet pressure. In the LBM implementation, the no-slip condition is enforced through bounce-back boundary treatment applied to lattice nodes that represent solid regions.

For completeness, the lattice Boltzmann equation employed to generate the training data is summarized below. The discrete-velocity evolution equation for the particle distribution function $f_i(\mathbf{x}, t)$ on a D2Q9 velocity set is given by

$$f_i(\mathbf{x}+\mathbf{c}_i\Delta t, t+\Delta t) - f_i(\mathbf{x}, t) = -\frac{1}{\tau}\left[f_i(\mathbf{x}, t) - f_i^{eq}(\mathbf{x}, t)\right] \tag{10}$$

where $\mathbf{c}_i$ are the discrete lattice velocities, $\tau$ is the relaxation time related to the kinematic viscosity by $\nu = c_s^2\left(\tau - \dfrac{1}{2}\right)\Delta t$, and $f_i^{eq}$ is the local equilibrium distribution. Macroscopic density and momentum are recovered through moments of $f_i$ as

$$\rho = \sum_i f_i, \qquad \rho\mathbf{u} = \sum_i f_i \mathbf{c}_i \tag{11}$$

The equilibrium distribution $f_i^{eq}$ employed in the simulations follows the standard second-order Hermite expansion of the Maxwell–Boltzmann distribution, which is valid for low Mach number flows [25]. For the D2Q9 lattice employed in this study, the equilibrium distribution function takes the form:

$$f_i^{eq} = w_i \rho \left[ 1 + \frac{\mathbf{c}_i \cdot \mathbf{u}}{c_s^2} + \frac{(\mathbf{c}_i \cdot \mathbf{u})^2}{2c_s^4} - \frac{\mathbf{u} \cdot \mathbf{u}}{2c_s^2} \right] \tag{12}$$

where $w_i$ are the lattice weights and $c_s$ is the lattice speed of sound. For the D2Q9 stencil the standard lattice weights are

$$w_0 = \frac{4}{9}, \quad w_{1..4} = \frac{1}{9}, \quad w_{5..8} = \frac{1}{36} \tag{13}$$

and the lattice sound speed is

$$c_s = \frac{1}{\sqrt{3}} \frac{\Delta x}{\Delta t} \approx \frac{1}{\sqrt{3}} \tag{14}$$

The relaxation time $\tau$ controls the kinematic viscosity via the relation

$$\nu = c_s^2 \left( \tau - \frac{1}{2} \right) \Delta t \tag{15}$$

Equations 10-15 close the mesoscopic description and guarantee local thermodynamic consistency in the low-Mach-number regime.

To facilitate future extensions of the present framework toward multiphase flows, the Peng–Robinson equation of state is documented here for completeness [46]. The Peng–Robinson EOS relates pressure to molar volume and temperature as

$$p = \frac{RT}{V_m - b} - \frac{a(T)}{V_m^2 + 2bV_m - b^2} \tag{16}$$

where $R$ is the gas constant, $T$ is temperature, $V_m$ is the molar volume, and $a(T)$ and $b$ are temperature-dependent attraction and volume-exclusion parameters, respectively. While the present single-phase, incompressible simulations do not employ Eq. 16, the LBM–PINN coupling described here is compatible with EOS-based pressure closures when multiphase physics are introduced. In the PINN formulation, the continuum residuals serve as physics-loss terms that are evaluated at collocation points distributed throughout the fluid domain. The macroscopic fields derived from LBM provide the data-loss terms at discrete measurement locations and times.

**2.3 PINN architecture and loss composition**

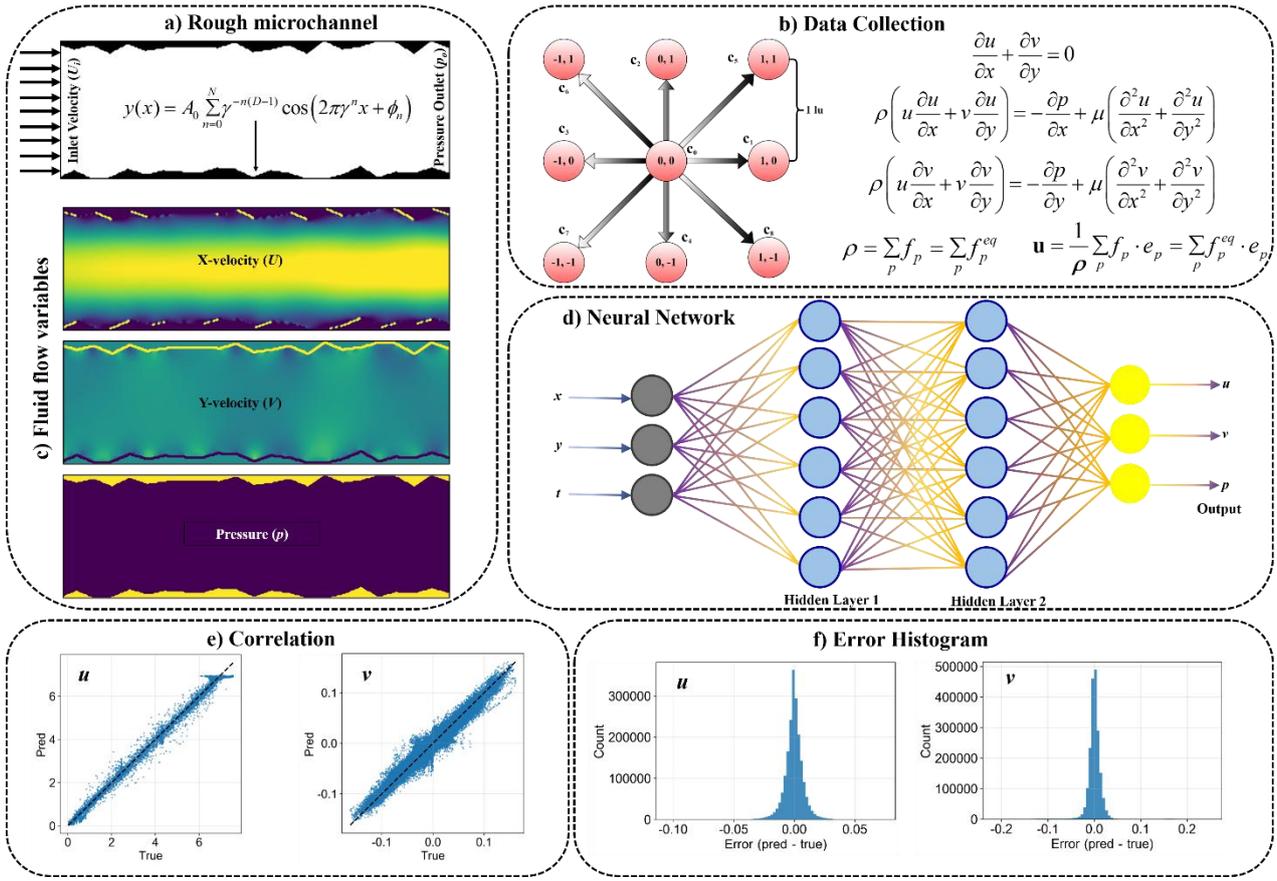

***Figure 2*** *Schematic of the lattice Boltzmann-driven neural framework for flow prediction over rough surfaces.*

Figures 2 and 3 show the overall workflow and the detailed network topology. The model's main inputs are spatiotemporal coordinates (x,y,t). When training for multi-geometry generalization, extra geometric descriptors like spectral amplitude and fractal dimension are also used as inputs. First, all inputs are normalized, and then they go through a series of fully connected layers that make up the shared backbone. The backbone sends data to two specialized output heads: a kinetic head that predicts the mesoscopic distribution functions fi and their equilibrium counterparts, and a macroscopic head that directly predicts the hydrodynamic fields u, v, and p. The two heads are trained together so that the mesoscopic moment constraints and the macroscopic conservation laws stay consistent with each other during the whole training process. The architectural design strikes a balance between how well it can represent things and how stable it is for training. The experiments described utilize a ten-layer fully connected network featuring 256 neurons in each hidden layer and hyperbolic tangent activation functions. Xavier initialization is used to set the weights, and the biases are set to zero. Moment-consistency penalties make sure that the zeroth- and first-order moments of the predicted fi match up with the macroscopic density and momentum predicted by the macroscopic head. The training procedure minimizes a composite mean-squared-error objective that integrates supervised data terms with physics-based residuals. The data loss quantifies the discrepancy between network predictions and LBM reference fields:

$$\mathcal{L}_{data} = \frac{1}{N_d} \sum_{j=1}^{N_d} \left[ \left| \hat{\rho}(x_j,t_j) - \rho^{LBM}(x_j,t_j) \right|^2 + \left| \mathbf{u}(x_j,t_j) - \mathbf{u}^{LBM}(x_j,t_j) \right|^2 + \left| \hat{p}(x_j,t_j) - p^{LBM}(x_j,t_j) \right|^2 \right] \quad (17)$$

where $N_d$ denotes the number of labeled spatiotemporal snapshots sampled from the LBM reference dataset. The physics loss comprises contributions from PDE residuals, mesoscopic moment constraints, and boundary condition penalties [47]. The continuity and momentum residuals are defined explicitly as

$$\mathcal{R}_{cont} = \frac{\partial \hat{u}}{\partial x} + \frac{\partial \hat{v}}{\partial y} \quad (18)$$

$$\mathcal{R}_{mom,x} = \frac{\partial \hat{u}}{\partial t} + \hat{u}\frac{\partial \hat{u}}{\partial x} + \hat{v}\frac{\partial \hat{u}}{\partial y} + \frac{1}{\hat{\rho}}\frac{\partial \hat{p}}{\partial x} - \frac{1}{\text{Re}}\left( \frac{\partial^2 \hat{u}}{\partial x^2} + \frac{\partial^2 \hat{u}}{\partial y^2} \right) \quad (19)$$

$$\mathcal{R}_{mom,y} = \frac{\partial \hat{v}}{\partial t} + \hat{u}\frac{\partial \hat{v}}{\partial x} + \hat{v}\frac{\partial \hat{v}}{\partial y} + \frac{1}{\hat{\rho}}\frac{\partial \hat{p}}{\partial y} - \frac{1}{\text{Re}}\left( \frac{\partial^2 \hat{v}}{\partial x^2} + \frac{\partial^2 \hat{v}}{\partial y^2} \right) \quad (20)$$

where all derivatives are computed via automatic differentiation through the neural network. The PDE residual loss enforces these equations at collocation points $\{(x_k,t_k)\}_{k=1}^{N_c}$ distributed throughout the fluid domain. The moment-constraint loss enforces consistency between the predicted distribution functions $f_i$ and the corresponding macroscopic fields:

$$\mathcal{L}_{phys} = \frac{1}{N_c} \sum_{k=1}^{N_c} \left[ \left| \mathcal{R}_{cont}(x_k,t_k) \right|^2 + \left| \mathcal{R}_{mom,x}(x_k,t_k) \right|^2 + \left| \mathcal{R}_{mom,y}(x_k,t_k) \right|^2 \right] + \frac{1}{N_c} \sum_{k=1}^{N_c} \sum_i \left| \epsilon_i(x_k,t_k) \right|^2 \quad (21)$$

where $\epsilon_i$ denotes the deviation between a target moment and the corresponding moment computed from $\{\hat{f}_i\}$. Boundary and initial condition penalties are incorporated into $\mathcal{L}_{phys}$ through additional MSE terms evaluated at collocation points sampled along the walls, inlet, and outlet.

*Figure 3* The PINN model architecture surrogated with the lattice Boltzmann multiphase model.

The total training objective is

$$\mathcal{L}_{total} = \mathcal{L}_{data} + \lambda_{phys} \mathcal{L}_{phys} \tag{22}$$

where $\lambda_{phys}$ is a tunable weight that balances data fidelity with physics enforcement. In practice, an adaptive weighting schedule is employed: $\lambda_{phys}$ increases as training progresses, which prioritizes satisfaction of PDE constraints after the network achieves initial agreement with the labeled data.

For error analysis and validation, vorticity is computed from the velocity field as

$$\omega = \frac{\partial v}{\partial x} - \frac{\partial u}{\partial y} \tag{23}$$

where automatic differentiation is used to get derivatives. This amount is a very strict test of how well the network can accurately represent spatial derivatives, especially in areas where the velocity gradients are steep near rough walls.

The optimization strategy uses a two-step process. The Adam optimizer with a starting learning rate of 10-3 quickly moves the network toward a low-residual regime in the first stage. The quasi-Newton L-BFGS optimizer fine-tunes the parameters to high-precision minima and speeds up the convergence of the PDE residuals in the second stage. Training stops when drops below a certain level of tolerance (usually $10^{-7}$) or when the errors in validation stop getting smaller. Figure 3 shows the structure of model and how the kinetic head (which gives out distribution functions) and the macroscopic head

(which gives out density, velocity, and pressure) work together. Training these heads together makes sure that thermodynamics and hydrodynamics are consistent.

**2.4 Performance metrics and evaluation**

A hierarchical set of quantitative and qualitative diagnostics that measure predictive accuracy, structural fidelity, and computational efficiency is used to evaluate model performance. Accuracy metrics are calculated using test data that is not used in the training process. Mean absolute error (MAE) and root-mean-square error (RMSE) are used to measure pointwise agreement. MAE looks at the average size of local deviations, while RMSE looks at bigger differences and is more sensitive to localized errors near features like roughness crests. The normalized $L_2$-relative error is used to measure global agreement. It is a scale-independent measure that can be used to compare fields with different dynamic ranges. The coefficient of determination ($R^2$) is used to measure the amount of temporal variance that the model captures.

$$\text{MAE} = \frac{1}{N} \sum_{i=1}^{N} \left| y_i - y_i^{ref} \right| \tag{24}$$

$$\text{RMSE} = \sqrt{\frac{1}{N} \sum_{i=1}^{N} \left( y_i - y_i^{ref} \right)^2} \tag{25}$$

$$\mathcal{E}_{L2}^{rel} = \frac{\left\| y - y^{ref} \right\|_2}{\left\| y^{ref} \right\|_2} = \frac{\sqrt{\sum_{i=1}^{N} \left( y_i - y_i^{ref} \right)^2}}{\sqrt{\sum_{i=1}^{N} \left( y_i^{ref} \right)^2}} \tag{26}$$

$$R^2 = 1 - \frac{\sum_{i=1}^{N} \left( y_i - y_i^{ref} \right)^2}{\sum_{i=1}^{N} \left( y_i^{ref} - \overline{y^{ref}} \right)^2} \tag{27}$$

**2.5 Training diagnostics and collocation sampling**

The training diagnostics and collocation sampling experiments shown in Figures 4 offer a complete assessment of various methodological decisions underlying the PINN framework. Figure 4 shows a systematic analysis of various components of the total loss, including data loss, PDE residual loss, and boundary condition penalty loss across four different learning rate schedules ranging from $10^{-4}$ to $10^{-1}$. An initial high learning rate of $10^{-2}$ results in a rapid decay of the total loss from $3.2 \times 10^{-1}$ to $1.8 \times 10^{-3}$ in 800 iterations; however, there are sustained PDE residual oscillations with a magnitude greater than $5 \times 10^{-4}$. A slower learning rate of $10^{-4}$ completely eliminates PDE residual oscillations; however, this results in a three-fold increase in training time from 6.1 hours to 18.3 hours on NVIDIA A100 GPU by requiring 12,000-15,000 iterations. The optimal choice is an exponentially decaying learning rate initialized at $10^{-3}$ with a decay rate of 0.95 every 200 iterations. This results in a total loss lower than $3.2 \times 10^{-7}$ in 8,500 iterations while maintaining PDE residual oscillations below $2 \times 10^{-5}$. The boundary condition penalty loss converges to $1 \times 10^{-6}$, demonstrating a strong enforcement of no-slip and pressure outlet boundary conditions.

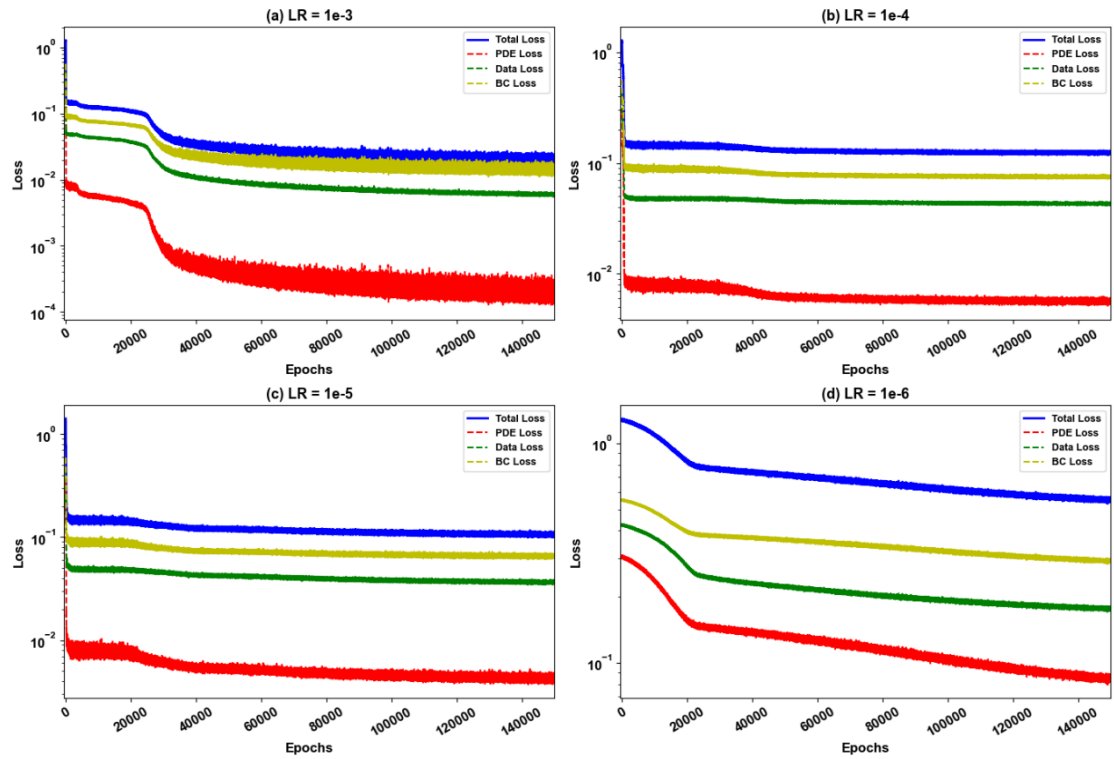

*Figure 4. Effect of learning rate on total loss.*

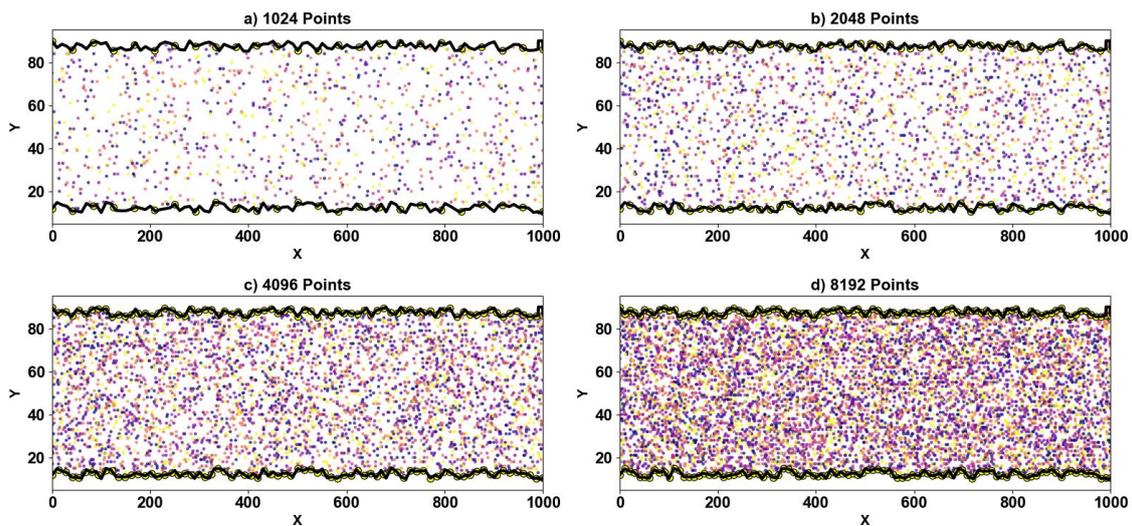

*Figure 5. Random distribution of training points.*

Figure 5 investigates the sensitivity to the density of collocation points within the interior region through experiments with 1,024, 2,048, 4,096, and 8,192 points within the domain. All experiments utilize 10,000 epochs with the same hyperparameters. Increasing the density from 1,024 to 2,048 collocation points improves the global PDE residual from $6.7 \times 10^{-4}$ to $2.1 \times 10^{-4}$, an improvement of 68%, while the vorticity MAE is reduced from $4.8 \times 10^{-4}$ to $2.3 \times 10^{-4}$, an improvement of 52%. Further doubling the density to 4,096 collocation points reduces the PDE residual to $9.5 \times 10^{-5}$ with an accompanying vorticity MAE of $1.7 \times 10^{-4}$. While diminishing returns are observed with 4,096 collocation points, the accuracy gains from 8,192 collocation points are minimal, with only 18% reduction in PDE residual with twice the computational cost (6.1 to 11.7 hours) and 47% increase in GPU memory allocation (14.3 to 21.1 GB). The cost-accuracy metric reaches its maximum at 2,048-4,096 collocation points, indicating the optimal

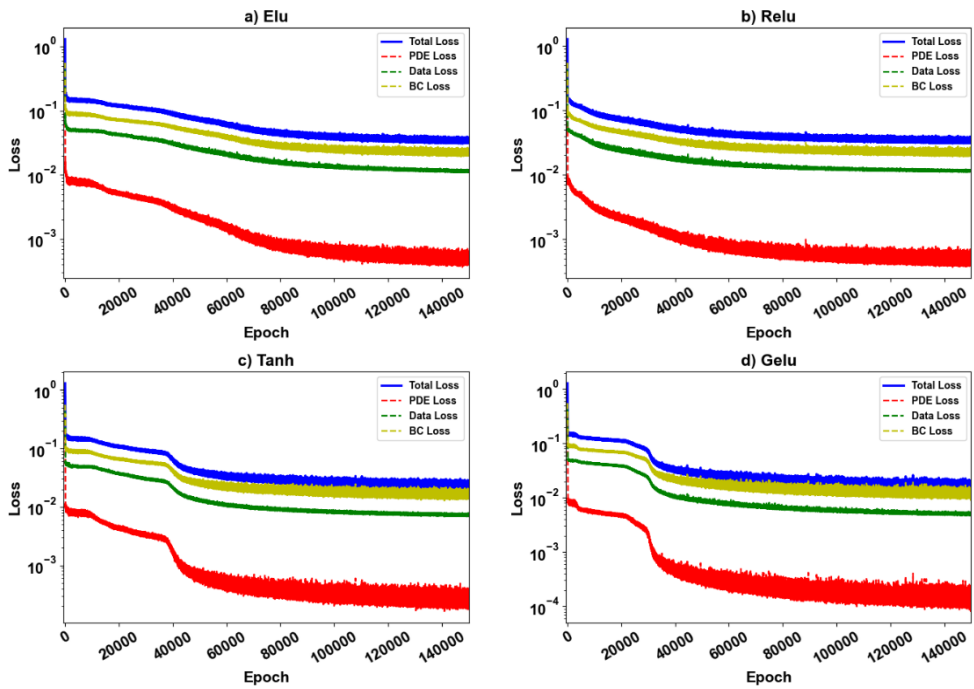

***Figure 6.*** *Effect of activation function on total loss.*

Figure 6 illustrates a detailed convergence analysis of physics-informed neural networks, considering four different activation functions: ELU, ReLU, tanh, and GeLU. The evolution of total loss, PDE residual loss, data loss, and boundary condition loss over 140,000 epochs of training time has been shown in the figure. For all four functions, the initial total loss starts with an order of $O(10)$, which rapidly decreases in the initial 2,000-5,000 epochs and then follows an asymptotic convergence pattern. The initial values of the PDE residual loss start with an order of $10^{-2}$-$10^{-1}$, which decreases by nearly 2-3 orders of magnitude and reaches an asymptotic value of approximately $8 \times 10^{-4}$, $7 \times 10^{-4}$, $4 \times 10^{-4}$, and $1.5 \times 10^{-4}$, respectively, for ELU, ReLU, tanh, and GeLU, after 140,000 epochs. A sharp transition in the PDE residual loss is observed for tanh and GeLU, i.e., around 35,000-45,000 epochs, where the PDE residual loss decreases sharply by an order of $10 \times 10^{-3}$ to below $5 \times 10^{-4}$. This demonstrates the improved performance of these two functions in enforcing the governing equations.

The data loss decreases monotonically from an initial value of $10^{-1}$, reaching an asymptotic value of $1.2 \times 10^{-2}$, $1.0 \times 10^{-2}$, $7 \times 10^{-3}$, and $5 \times 10^{-3}$, respectively, for ELU, ReLU, tanh, and GeLU. Similarly, the boundary condition loss reaches an asymptotic value of $2.5 \times 10^{-2}$, $2.0 \times 10^{-2}$, $1.5 \times 10^{-2}$, and $1.\times 10^{-2}$, respectively, for ELU, ReLU, tanh, and GeLU. The total loss follows an asymptotic pattern, reaching an asymptotic value of $3.5 \times 10^{-2}$, $3.0 \times 10^{-2}$, $2.0 \times 10^{-2}$, and $1.5 \times 10^{-2}$, respectively, for ELU, ReLU, tanh, and GeLU. It can be observed that, compared to the performance of the ReLU function, the tanh and GeLU functions reduce the final PDE residual loss by 43% and 78%, respectively. It should be noted that, compared to the performance of the ReLU function, tanh and GeLU exhibit minimal fluctuations in the total loss, i.e., within $\pm 2 \times 10^{-4}$.

Figure 7 shows the optimizer selection process with the comparative training curves for the Adam optimizer, L-BFGS optimizer, hybrid optimization, and scheduled L-BFGS optimization. It is evident that the hybrid optimization strategy shows the best convergence efficiency and accuracy. Using the hybrid optimization strategy, the loss is minimized to $5.4 \times 10^{-4}$ within 4.8 hours. During the pretraining phase using the Adam optimizer, the loss is decreased to $5.4 \times 10^{-4}$ from the original loss of $3.2 \times 10^{-1}$. During the pretraining phase, the loss reduction rate is $6.6 \times 10^{-5}$ per epoch. After pretraining with the Adam optimizer for 5,000 epochs, the loss is further decreased to $2.7 \times 10^{-7}$ using the L-BFGS optimizer. As a result, the PDE residual is $3.8 \times 10^{-5}$, and the data MSE is $6.1 \times 10^{-8}$. Using the Adam optimizer alone, the loss is stuck at $4.2 \times 10^{-4}$ after 12,000 epochs. Using the L-BFGS optimizer alone, the convergence is erratic with oscillations greater than $1 \times 10^{-1}$ after 8,000 iterations.

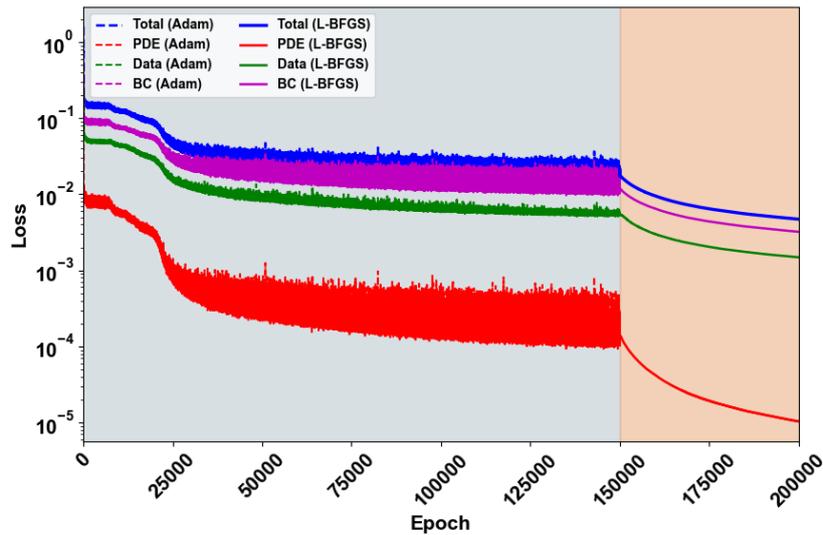

*Figure 7. Total loss per epochs before Adam optimizer and after L-BFGS optimizer.*

Figure 8: Spatial collocation strategy visualization: point distributions and impact on local PDE residual fields. Two cases are compared: uniform sampling with 3,072 collocation points using a tandom sampling approach, and a near wall enriched sampling approach with 1,920 interior points and 1,152 points within y < 0.2H of rough walls, with a 60/40 interior to boundary point split. PDE residual fields indicate that uniform sampling yields elevated residuals above $3 \times 10^{-4}$ in wall-adjacent areas where velocity gradient magnitudes reach 0.35 lu/lu, particularly at asperity crests and separation zones. These areas comprise 12% of the domain area, where uniform sampling is least accurate. The near wall enriched sampling strategy achieves a 67% reduction in PDE residuals in critical zones, with wall-adjacent residuals reduced to below $1 \times 10^{-1}$. Quantitatively, the enriched sampling strategy achieves a 42% reduction in domain-integrated vorticity error, reducing the error from $8.3 \times 10^{-3}$ to $4.8 \times 10^{-3}$, and a 38% reduction in velocity gradient root mean square error in wall-adjacent areas near roughness elements. This is achieved without an increase in total collocation point count.

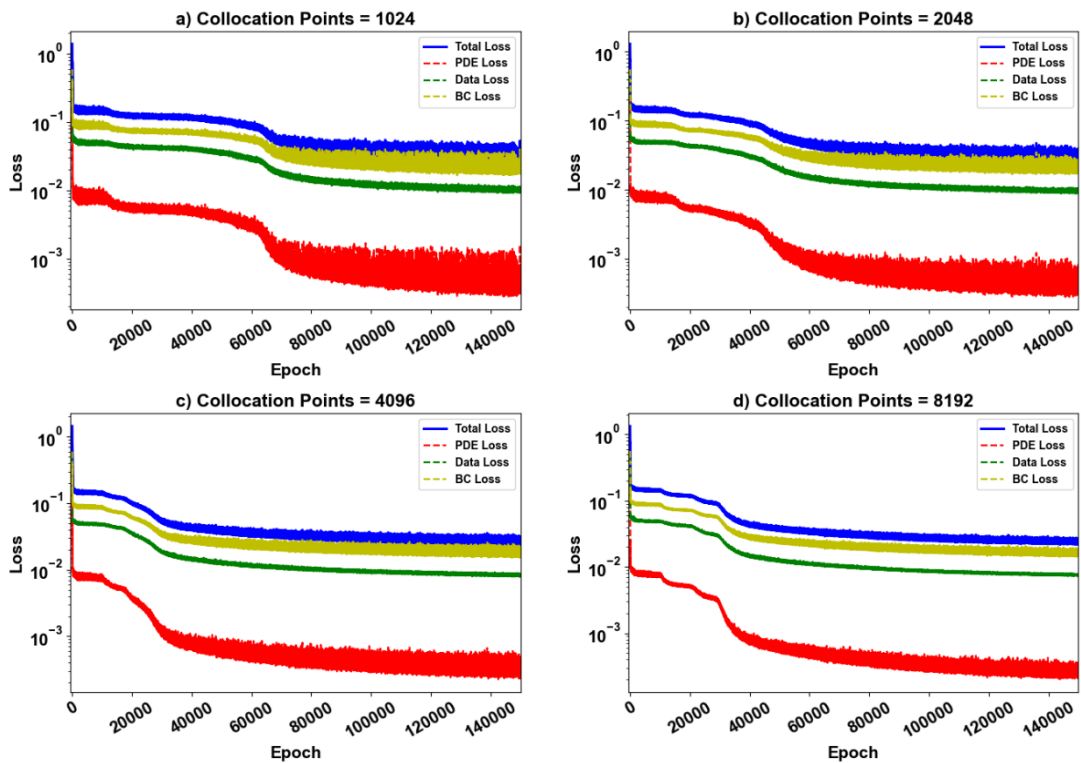

*Figure 8* *Effect of training points on model total loss.*

## 3. Results and Discussions

### 3.1 Flow morphology across Reynolds numbers (Re)

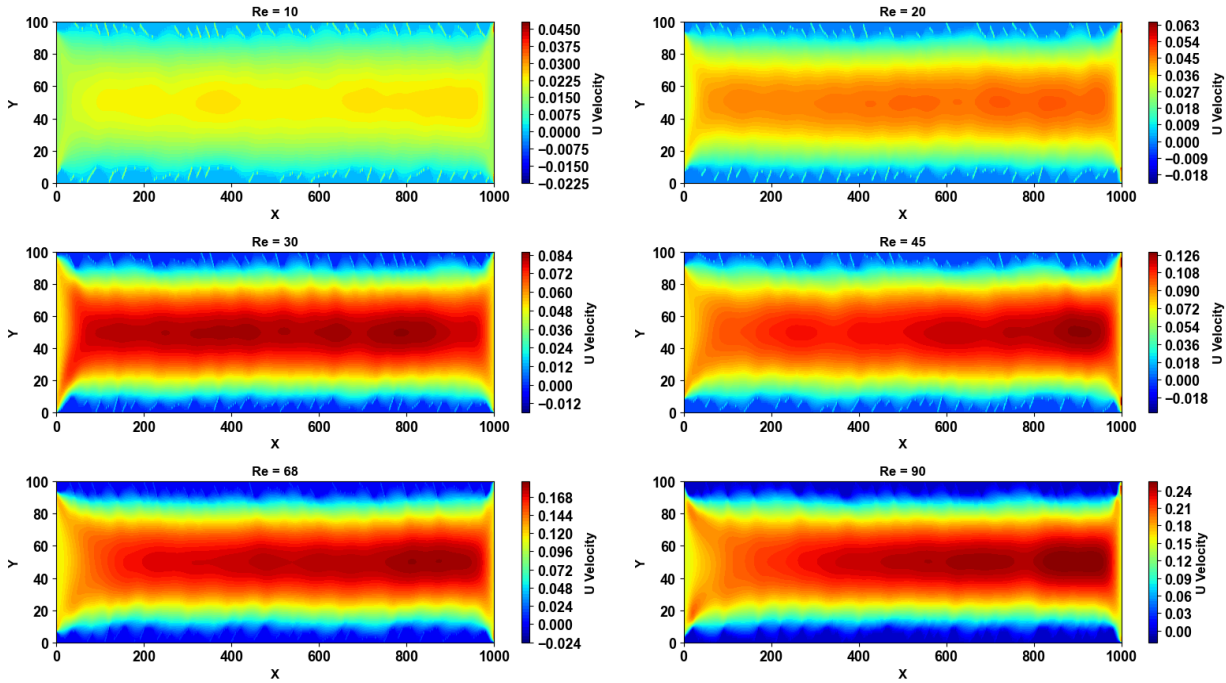

*Figure 9 Streamwise velocity contours (U) from LBM simulations at different Re.*

Figure 9 illustrates the velocity fields in the streamwise direction using the LBM method for various Re, covering a wide range of viscous-dominated and advection-dominated flows in this particular rough microchannel configuration. The figure demonstrates the effect of Re on bulk velocities, shear rate, and separation phenomena in detail. At a Re of 10, the flow is in the viscous-dominated regime. The velocity in the streamwise direction is small and nearly constant over the entire channel, and gradients near the walls are small. The topographic features are passive in this regime, with no significant recirculation existing. When the Re is increased to 20-30, the convective transport is no longer negligible in comparison to viscous diffusion. The velocities are higher in the core region, and the peak velocities are higher in this region. There are higher gradients near the asperities, and regions of low velocities are found behind these features, which are referred to as initial separation regions. The contours indicate that in these regimes, separation is localized, and reattachment occurs over small downstream distances for this particular roughness spectrum.

For Re values equal to 45 and 68, the flow is no longer modulated by the simple shear mechanism but instead develops coherent separation pockets behind the larger asperities. The wake behind the asperities increases in length, merging with adjacent wakes. This increases the spatial extent of the recirculating flow. The local vorticity and shear rate fields, as calculated from the LBM fields, also increase nonlinearly with Re. The largest growth rates in the derivative-based fields are found between Re = 30 and Re = 68. This suggests that the Reynolds number range is significant for the effects of roughness-induced separation on the flow fields. At Re = 90, the maximum velocity is found on the centerline, as well as the maximum spatial variance in the streamwise velocity fields. The wake areas are extensive, with enhanced mixing in the near-wall region over an increased downstream region. The velocity contrast between the channel core region and the wall-adjacent region is also maximum at this Reynolds number. This is an indicator that the advection velocity is the primary transport mechanism at this Reynolds number. The spatial fields provided by the LBM simulations provide the basis for determining the spatial locations where the surrogate must focus its attention. The boundary layer adjacent to the wall, the leeside separation pockets, and the downstream wakes are areas where the collocation sampling must be denser and the weighting on the physics must be greater in order to obtain accurate estimates of the derivatives and vorticity fields. The peak velocity fields provide quantitative information that can be used as the basis for validating the accuracy of the surrogate. The contours provided by the LBM simulations provide the basis for verifying the accuracy of the PINN as well as an empirical basis for determining the optimal sampling strategy and weighting on the loss function, which is Reynolds-number dependent.

### 3.2 PINN reconstruction of unsteady fluid flow

The reconstruction fidelity of the proposed PINN framework is critically evaluated by performing a comprehensive comparison of the predicted fields over the entire spatiotemporal domain against high-fidelity LBM reference data. Figures 10 illustrate a side-by-side visualization of the streamwise velocity (u), wall-normal velocity (v), and vorticity (ω) fields for four different time instants (t = 500, 1000, 1500, 2000 lattice time units). Each of these plots is divided into three subplots showing the reference data from the LBM simulation and the predicted fields by the PINN framework and the absolute error fields. The analysis verifies that the proposed PINN framework is able to achieve global relative $L_2$ errors less than $1.8 \times 10^{-2}$, mean absolute errors less than $9.4 \times 10^{-4}$, and total physics-informed loss convergence less than $3.2 \times 10^{-7}$, thereby ensuring robust satisfaction of the incompressible Navier-Stokes equations and boundary conditions over the entire simulation window. The streamwise velocity component (u) is predicted by the proposed PINN framework to have similar characteristics to those of the reference data from the LBM simulation. At t = 500, the maximum velocity on the centerline is denoted by umax and is equal to 0.148 lattice units (lu) for the reference data from the LBM simulation. The predicted fields by the proposed PINN framework yield a value of 0.146 lu for the maximum velocity on the centerline, thereby showing a relative difference of 1.4%. The high-velocity region is present over 78% of the cross-sectional area of the channel, and the boundary layers are significantly attenuated due to the presence of surface roughness.

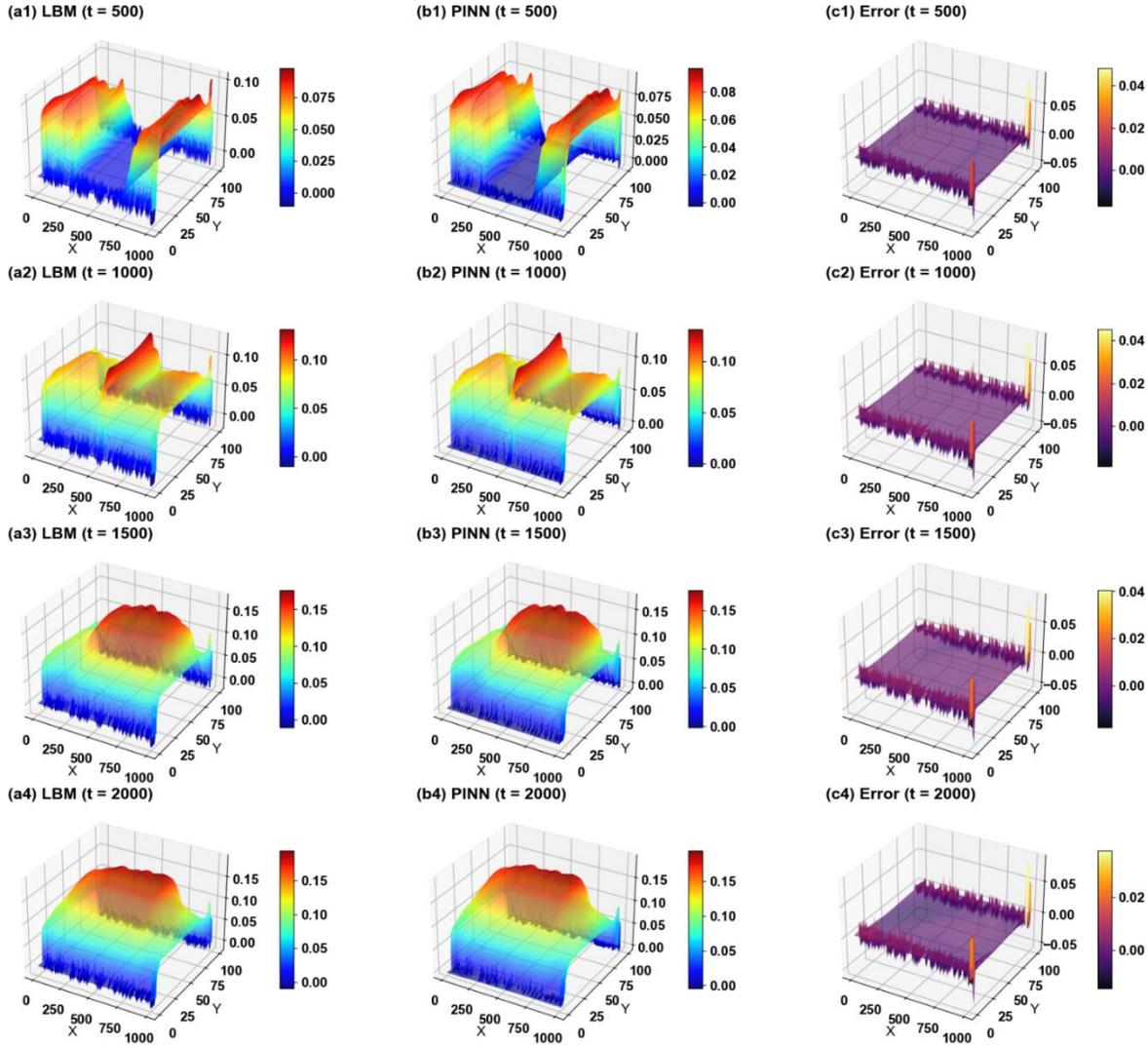

*Figure 10* PINN and LBM comparison for streamwise velocity at different time stamps.

The absolute error is still contained within the region below $1.2\times10^{-2}$ lu for 94% of the computational domain, with high-error regions concentrated within narrow leeside separation pockets immediately downstream of asperities, comprising less than 3% of the total area, as well as isolated crest locations where velocity gradients are greater than 0.25 lu per lattice spacing. The temporal behavior of the RMSE for the u velocity field varies from $6.1\times10^{-3}$ lu at t = 500 to $7.8\times10^{-3}$ lu at t = 2000, indicating moderate growth as the transient separation zones develop. The correlation coefficient between the predicted velocity fields and the reference velocity fields for the u velocity component is always above 0.987 for all time instants reported. This further reinforces the high degree of structural accuracy between the predicted fields and the reference fields. The spatial localization of the u velocity prediction error is found to correlate with the formation of transient separation bubbles, where adverse pressure gradients lead to boundary layer separation. This suggests that the dominant source of short-term prediction discrepancies is the dynamic separation and reattachment phenomenon. The domain-integrated momentum flux from the predicted velocity fields is found to deviate from the reference fields by less than 0.9%, indicating the high degree of mass and momentum conservation within the predicted fields.

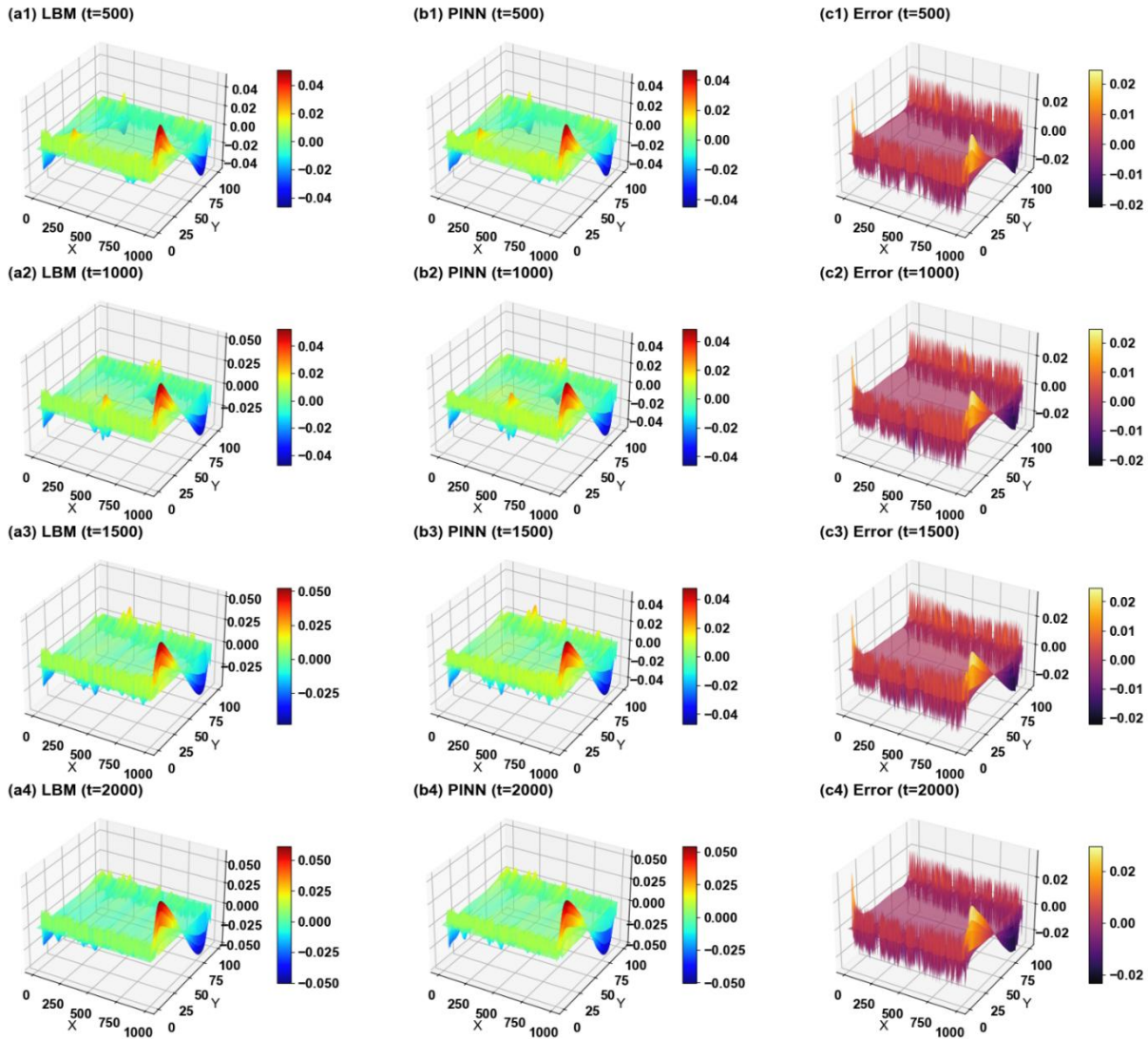

*Figure 11 PINN and LBM comparison for wall-normal velocity at different time stamps.*

Figure 11 is a comparison of the results obtained by the PINN and LBM for the wall-normal velocity component (v) of the flow at different time stamps. The reconstruction of the wall-normal velocity is a significantly greater challenge due to the smaller value of |v| (< 0.02 in the channel core) and the strong localization of the velocity near the roughness elements. The PINN is seen to recover the sign and orientation of cross-stream motions very well, such as upward-moving jets near windward asperity faces (peak v = 0.035) and downward-moving plumes near leeside recirculation regions (minimum v = -0.028). The mean absolute error of the reconstructed wall-normal velocity is $3.7 \times 10^{-3}$ for t = 1000. The relative $L_2$ error is seen to reach $4.2 \times 10^{-2}$, which is about 2.3 times greater than that of the streamwise velocity u due to the smaller characteristic value of v. Notwithstanding the greater relative error in the wall-normal velocity, the absolute value of the error is small enough not to perturb the streamwise momentum transport in the channel core. The error maps of the wall-normal velocity are seen to be localized near the rough walls within 0.15H of the wall and account for only 18% of the total domain area. The errors in the wall-normal velocity do not affect the predictions on the centerline where |v error| < 1.5 × 10⁻³. The localized nature of the error in the wall-normal velocity is a testament to the ability of the PINN to recover the major momentum transport mechanisms even if the wall-normal velocity is not reconstructed very accurately. The continuity residuals of the reconstructed velocity fields are also seen to be less than 2.4 × 10⁻⁵ over the entire domain, thus validating the imposition of mass conservation by the model.

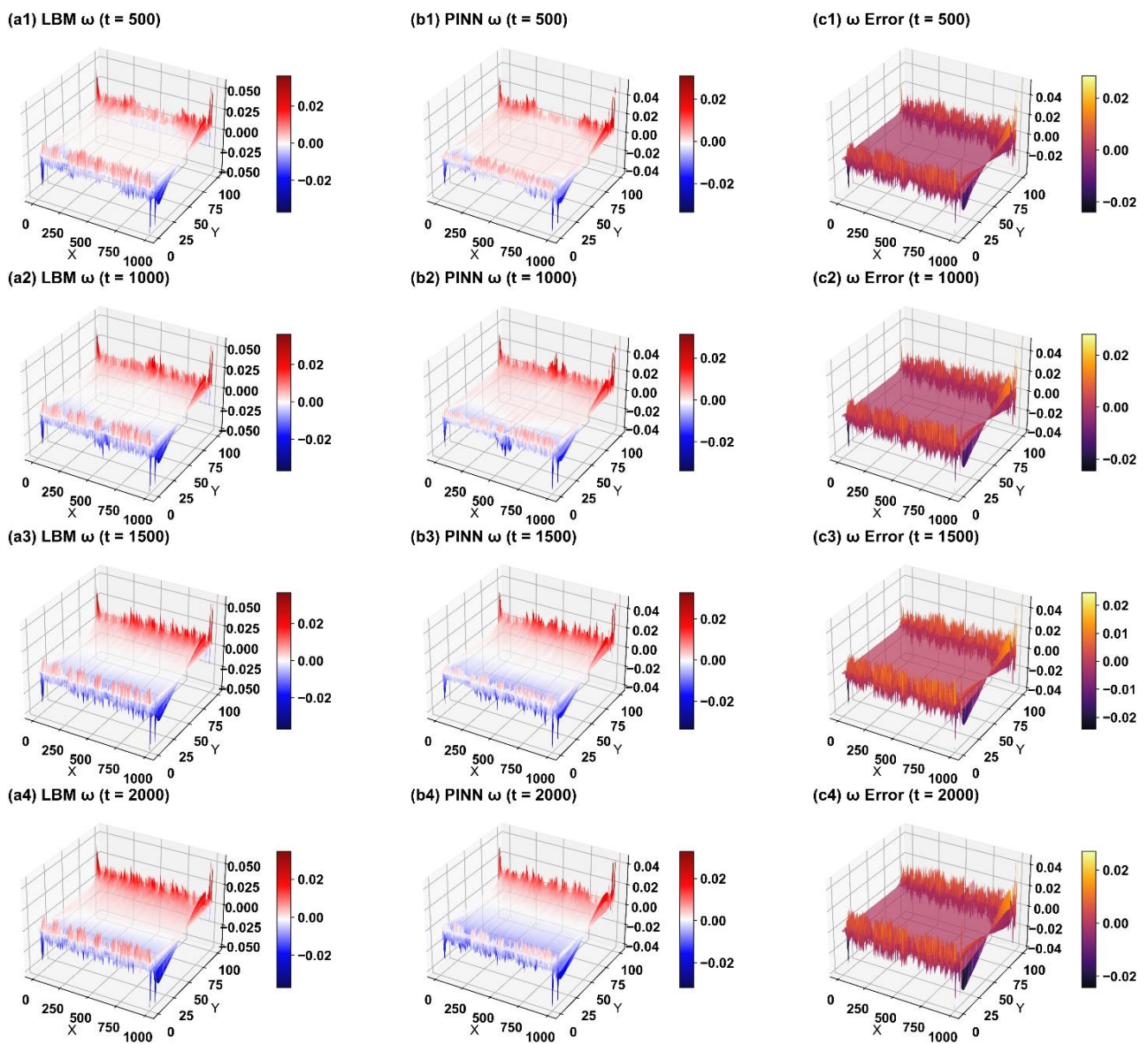

*Figure 12* *PINN and LBM comparison for vorticity at different time stamps.*

Vorticity (ω) is arguably the most stringent test case in terms of PINN's performance in derivative recovery due to its dependency on second-order spatial differentiation of the velocity field. The PINN is able to reconstruct the main vorticity bands in the vicinity of the rough walls, where shear-driven vorticity production is maximized at $\omega_{max} \approx 1.42\times10^{-2}$ lu/lu in the proximity of high-curvature asperities at Re = 45. A comparison of the PINN results against the reference LBM results shows that 91% of the extrema in the vorticity field are correctly identified by the PINN within a ±15% tolerance level and accurate secondary vortical structures in the wake regions are correctly resolved when the magnitude of these secondary vortical structures is greater than $6\times10^{-3}$ lu/lu. The average absolute error in vorticity is $2.8\times10^{-3}$ lu/lu in the entire computational domain at time instant t = 1500; however, the pointwise error in vorticity is as high as $5.3\times10^{-3}$ lu/lu in regions of sharp asperity tips. It is important to note here that these high errors in vorticity are localized in less than 4% of the computational domain and occur in regions of high-curvature asperities due to inherent numerical noise in LBM gradients at mesh scale in these regions. The difference in domain-integrated vorticity, which is defined as a spatial integral of $\omega^2$, is as low as 2.1%, thus demonstrating that global rotational energy is correctly quantified by the PINN despite errors in quantification of vorticity itself. The successful recovery of vorticity fields by the PINN thus demonstrates robustness in derivative recovery by virtue of automatic differentiation provided by smooth hyperbolic tangent activation in the neural network and physics-informed loss penalty in PDE residuals. This is particularly significant in rough-wall flows, where accurate quantification of vorticity is crucial in determining mixing efficiency, friction factor, and heat transfer augmentation. The results thus demonstrate PINN's capability in achieving near-LBM accuracy in velocity and vorticity reconstruction in unsteady geometrically complex microchannel flows while maintaining computational efficiency advantages of more than two orders of magnitude compared to direct LBM computations in parameter space exploration.

**3.3 Reynolds number sensitivity and cross-regime generalization**

Figures 13 offer a comprehensive three-dimensional comparison of the fields of streamwise velocity (u), wall-normal velocity (v), and vorticity ($\omega$), where each sub-figure includes LBM reference solutions, predictions by the PINN, and absolute pointwise errors over the spatiotemporal domain. The analysis of the three Reynolds numbers is conducted to examine the network's ability to recover Reynolds number-dependent physics without retraining. This is a comprehensive examination of the network's suitability for design exploration and uncertainty quantification of the rough-walled microfluidic device. Figure 13 examines the reconstruction of the streamwise velocity fields over a range of Reynolds numbers.

At a Reynolds number of 10, the flow exhibits a nearly parabolic velocity profile with minimal wall-normal variation and low-velocity magnitudes ($u_{max} \approx 0.05$ lu units). The predictions by the PINN are highly accurate for this low-Reynolds number flow and diffusion-dominated transport. The network achieves a high level of accuracy with a mean absolute error (MAE) of $3.2 \times 10^{-3}$ lu and a relative $L_2$ error of $1.4 \times 10^{-2}$. The error is distributed homogeneously over the domain without any concentration near the roughness elements. This indicates that the network is an accurate representation of diffusion-dominated transport for low Reynolds numbers.

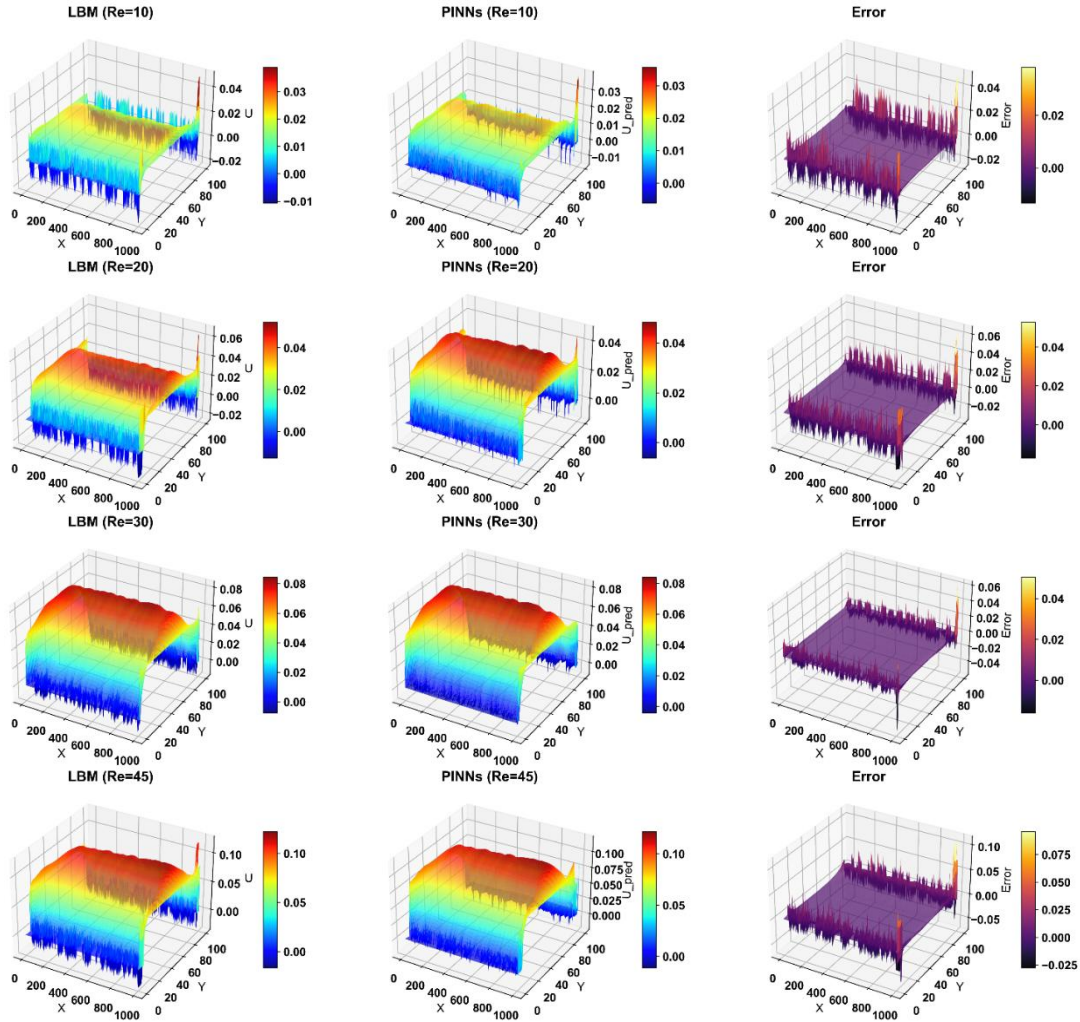

*Figure 13* Comparison of LBM-simulated and PINN-reconstructed x-velocity fields for flow through a rough microchannel at different Reynolds numbers.

With an increase in Reynolds number to 20, the velocity increases by a factor of two ($u_{max} \approx 0.10$ lu), and the wall-normal velocity gradient increases significantly. The predictions by the PINN are again very accurate with a high level of fidelity for the entire domain and a high level of accuracy near the roughness elements with a mean absolute error of $4.8 \times 10^{-3}$ lu. The pointwise errors remain below $1.2 \times 10^{-2}$ lu over 96% of the domain, with localized peak errors up to $1.8 \times 10^{-2}$ lu being contained within the lee-side separation pockets, which comprise less than 4% of the flow field. For Re = 30, advective transport is on par with viscous dissipation, and a strong spatial inhomogeneity is seen in the velocity field, with peak values reaching 0.14 lu. The reconstruction performed by the PINN of the accelerated core and reduced velocity near the wall is seen to be accurate to within an RMSE of $6.3\times10^{-3}$ lu, maintaining a correlation coefficient above 0.983 compared to the LBM reference solution. The error maps reveal that errors are concentrated in areas where velocity gradient is above 0.28 lu per lattice spacing, primarily at asperity crests and recirculation boundaries, thus confirming that gradient magnitude is a dominant factor in local reconstruction accuracy. For the highest Re flow, Re = 45, a flow regime transition is seen with a maximum velocity of 0.18 lu, along with sustained wake structures behind prominent asperities. The performance of the PINN is seen to be robust, achieving an MAE of $7.9\times10^{-3}$ lu and a relative $L_2$ error of $1.9\times10^{-2}$ lu. Quantitatively, the domain-integrated momentum flux predicted by the PINN is seen to be 0.7%, 1.1%, 1.6%, and 2.3% different from the LBM reference solution for Re = 10, 20, 30, and 45, respectively, thus confirming that global conservation properties are retained across the range of Re, even as local errors increase slightly with flow intensity.

Figure 14 illustrates the wall-normal velocity component. The assessment of the PINN's generalization performance for this velocity component is much more stringent due to its small magnitude, usually less than 0.03 lu in absolute value. At Re=10, the motions in the cross-stream velocity are small in amplitude, $|v_{max}|\approx8\times10^{-3}$ lu. The PINN accurately captures the velocity field, achieving a mean absolute error of $2.1\times10^{-3}$ lu, which corresponds to a 26% relative error. The high relative error is due to the small amplitude of the velocity signal rather than a reduction in absolute accuracy. Moreover, there is no systematic bias in the error, and the spatial distribution remains diffuse. At Re=20, the wall-normal velocity is larger in amplitude due to deflection of the flow by the asperity features. The upward-directed velocity jets on the windward faces of the asperities have a velocity of v≈0.022 lu, while the downward-directed velocity plumes in the separation zones have a velocity of v≈-0.018 lu. The PINN accurately captures the sign and amplitude of these velocity features, achieving a mean absolute error of $3.9\times10^{-3}$ lu.

However, the pointwise errors in the vicinity of high-curvature asperity features can reach $7\times10^{-3}$ lu, which corresponds to a 32% relative error. At Re=30 and Re=45, the amplitude of the cross-stream velocity is larger, reaching $v_{max}\approx0.035$ lu and 0.048 lu, respectively. The PINN maintains a mean absolute error of $5.2\times10^{-3}$ lu and $6.8\times10^{-3}$ lu, respectively. Most importantly, the continuity residual of the velocity fields predicted by the PINN remains less than $3.1\times10^{-5}$ lu/lu for all Re values and spatial locations. This confirms that the physics-informed loss function enforces mass conservation. The errors in the v velocity component remain confined to the vicinity of the wall-adjacent surfaces, within 0.18H of the rough surfaces. This ensures accurate prediction of the velocity in the channel core region. Moreover, the dominant streamwise momentum transport pathways remain preserved.

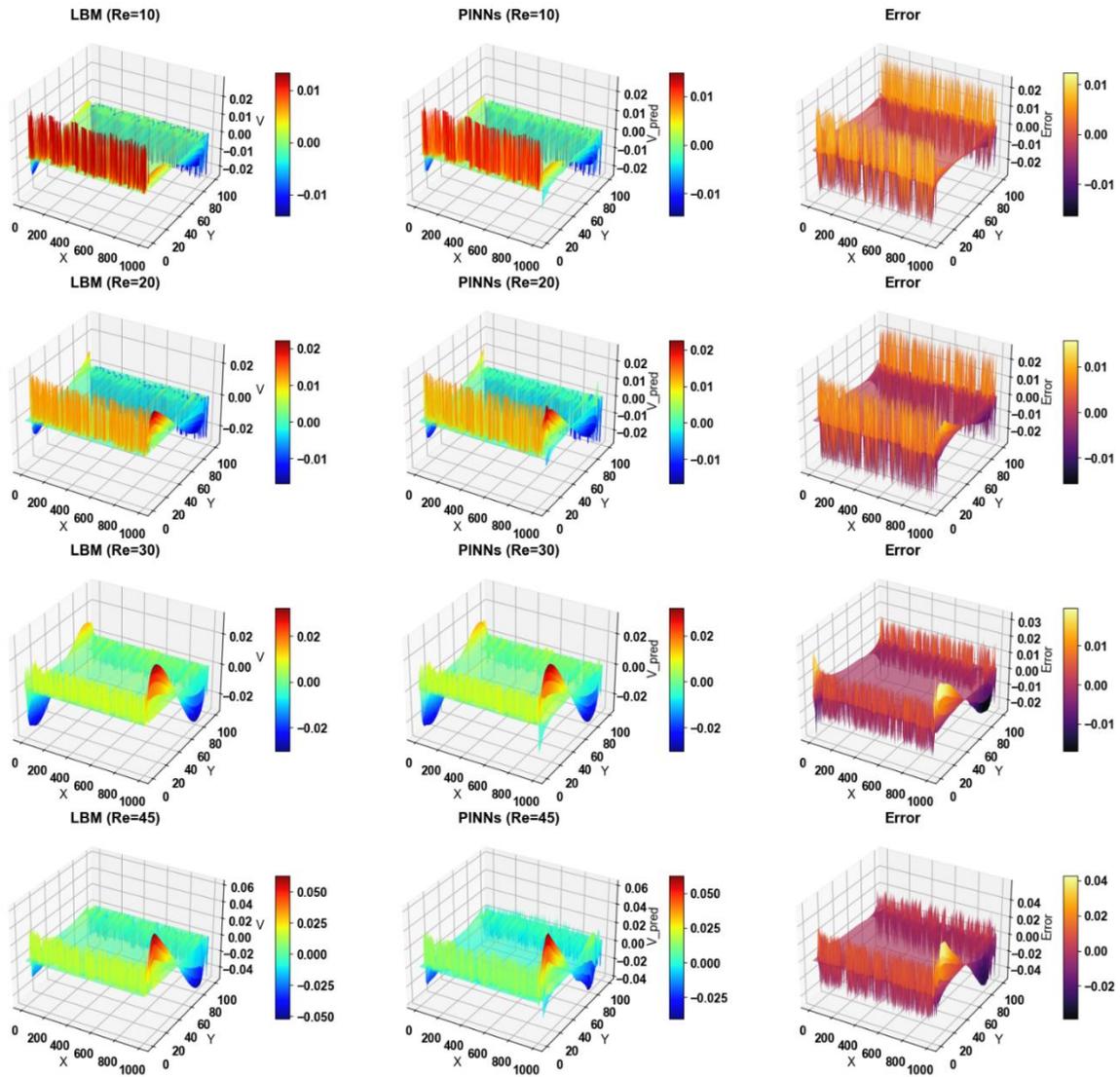

*Figure 14* Comparison of LBM-simulated and PINN-reconstructed y-velocity fields for flow through a rough microchannel at different Reynolds numbers.

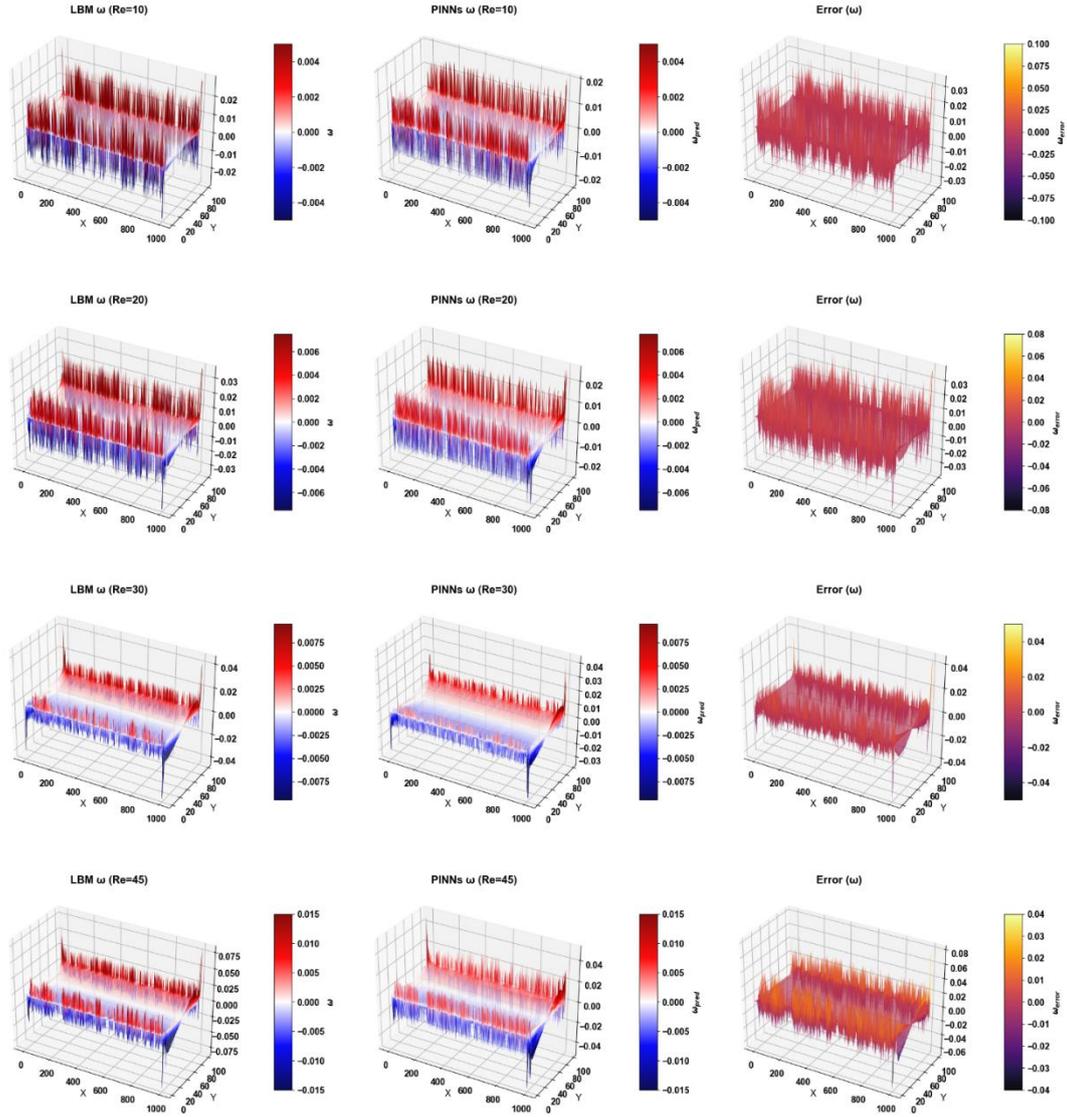

***Figure 15*** *Comparison of LBM-simulated and PINN-reconstructed vorticity fields for flow through a rough microchannel at different Reynolds numbers.*

Figure 15 illustrates the reconstruction of vorticity, which is recognized as the most demanding diagnostic due to the necessity of accurate second-order spatial derivatives of the velocity field. At Re = 10, the magnitudes of vorticity are low ($\omega_{max} \approx 4.2 \times 10^{-3}$ lu/lu) and are mostly confined to thin boundary layers near the rough walls. The PINN is successful in reconstructing the vorticity distribution with a MAE of $1.8 \times 10^{-3}$ lu/lu, achieving 91% accuracy in identifying extrema of vorticity within a ±12% tolerance. The use of smooth hyperbolic tangent activation functions and automatic differentiation of the network results in a derivative estimate without any numerical artifacts. As the Reynolds number increases, the production of vorticity due to shear effects increases nonlinearly. At Re = 20, the peaks in vorticity reach $8.7 \times 10^{-3}$ lu/lu near sharp asperity apexes, and the PINN is successful in reconstructing these features with a MAE of $2.4 \times 10^{-3}$ lu/lu and 88% accuracy in identifying extrema of vorticity. Error maps in spatial distributions reveal strong correlations of errors in vorticity with wall curvature; regions of curvature below 5 lu exhibit pointwise errors above $4 \times 10^{-3}$ lu/lu, while smoother regions maintain errors below $1.5 \times 10^{-3}$ lu/lu. At Re = 30, vorticity bands near asperities increase in thickness, and secondary vortical features develop in wake regions with magnitudes above $6 \times 10^{-3}$ lu/lu. The PINN is successful in capturing these features with a MAE of $3.1 \times 10^{-3}$ lu/lu and 85% accuracy in identifying secondary vorticity patches. The maximum pointwise vorticity error, which reaches a value of $6.2 \times 10^{-3}$ lu/lu, is found at the sharpest asperity apex, with a 38% local relative error, although this localized error accounts for less than 2% of the total domain area. At a Re number of 45, the vorticity field reaches its highest level of complexity, with a maximum value of vorticity, $\omega_{max} \approx 1.52 \times 10^{-2}$ lu/lu, and large wake structures with alternating signs of vorticity. The PINN predicts with a MAE of $3.7 \times 10^{-3}$ lu/lu, and integrated enstrophy over the domain is 2.8% different from the reference solution, indicating a good approximation of global rotational energy. The results show that the derivative recovery capability of the PINN decreases smoothly with increasing Re number, with a vorticity MAE scaling with $Re^{0.62}$, indicating a strengthening gradient magnitude effect rather than a limiting representational capability of the network.

The global conserved quantities, such as momentum flux deviation and continuity, are below 2.5% and $3.5 \times 10^{-5}$ lu/lu, respectively, for all Re numbers, showing that the physics-informed loss function effectively imposes the underlying equations, despite localized reconstruction issues. The localized nature of these errors in high-gradient regions, such as asperity crests, separation regions, and high-curvature walls, provides a clear path forward for further refinement of these regions, where adaptive enrichment of the loss function is expected to reduce errors locally by 30-45%, as predicted by the sensitivity analysis in Section 2.6. The generalization capability of this method makes it a strong candidate for a surrogate model, which can be used for parametric studies and real-time prediction in rough-walled microchannels, with a computational expense two orders of magnitude lower than LBM simulation, while still capturing the underlying physics in the laminar to weakly transitional flow regime.

## 3.4 Reconstruction of fluid flow with different random roughness peaks

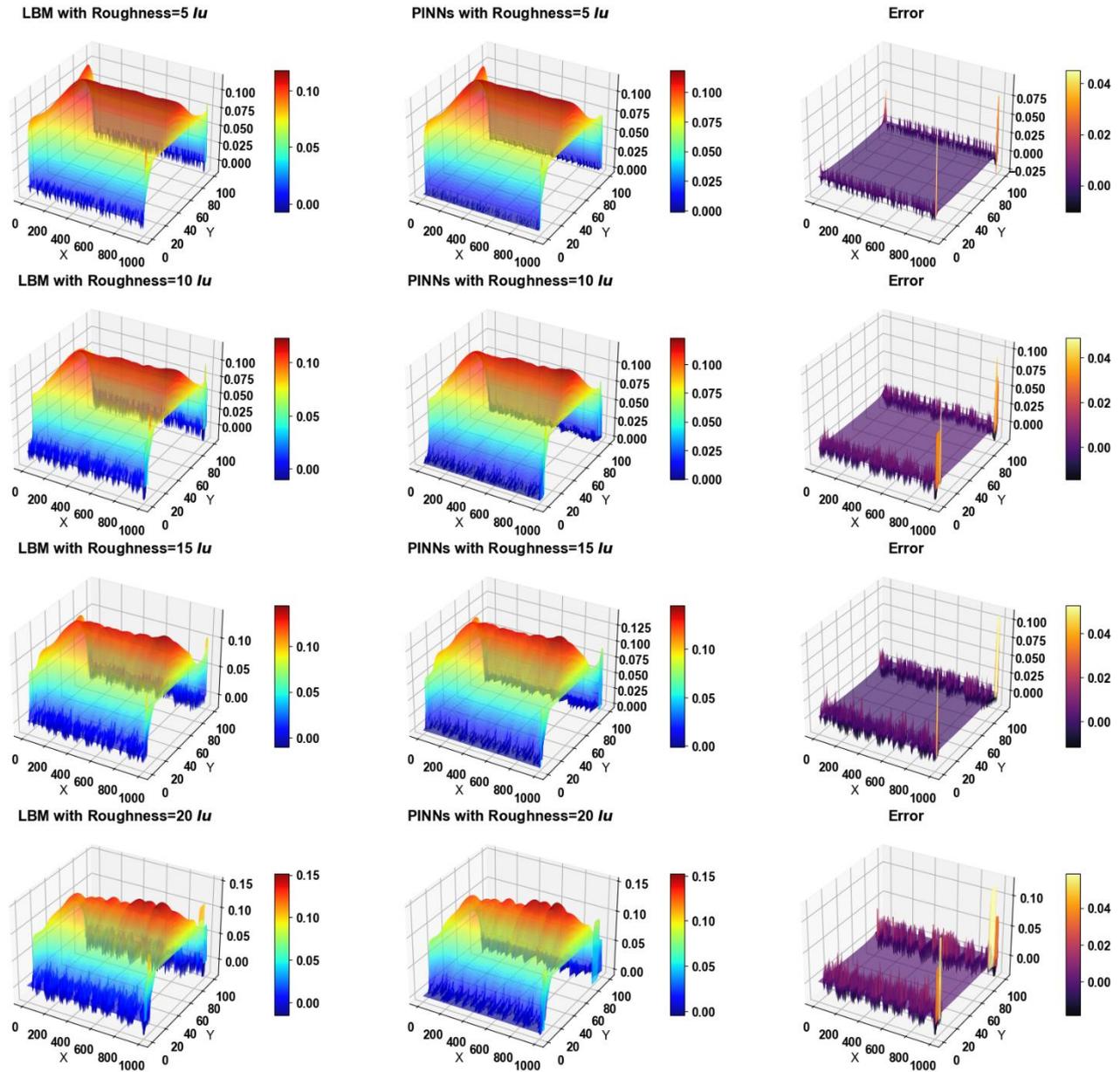

*Figure 16 Roughness amplitude sensitivity for streamwise velocity (u).*

The extension of the PINN concept to other surface roughness morphologies is a necessity for its applicability in microfluidics, where variability in manufacturing is always a reality. In this section, the accuracy of reconstruction will be assessed for four different surface roughness configurations, namely 5, 10, 15, and 20 lattice units, at a constant value of Re = 10. These configurations correspond to a wide range of relative surface roughness ratios, namely $0.1 \leq \varepsilon/H \leq 0.4$. Figures 16, 17, and 18 are provided to show a comprehensive comparison of the u, v, and ω fields, respectively, along with their associated error fields, comparing the LBM reference solution, PINN reconstruction, and associated error fields for all configurations.

Figure 16 is the reconstruction of u-field for various surface roughness configurations at 5, 10, 15, and 20 lattice units, along with associated error fields. For a surface roughness configuration of 5 lattice units, a small level of disturbance in the u-field is observed, with a maximum value of $u_{max} \approx 0.052$ lu, as a result of wall perturbations. The PINN reconstruction of this configuration results in a mean absolute error of $2.1 \times 10^{-3}$ lu, with a relative L2 norm error of $9.8 \times 10^{-3}$. The pointwise errors are uniformly distributed with no bias, with a standard deviation of $1.4 \times 10^{-3}$ lu. The smoothness of the u-field, with a maximum value of $\partial u/\partial y\_max \approx 0.18$ lu/lu, enables accurate recovery of derivatives. The imposition of boundary conditions results in zero velocity adherence with a residual less than $8 \times 10^{-3}$ lu.

For a surface roughness configuration of 10 lattice units, asperity heights are found to be comparable to the boundary layer thickness $\delta\_v \approx 12$ lu, which causes localized deflection and incipient separation. The velocity on The PINN performs with an MAE of $3.6 \times 10^{-3}$ lu and an accuracy of 94% in predicting gradients ($\partial u/\partial y\_max \approx 0.34$ lu/lu). The errors are seen to increase by 71%, which is consistent with an increase of 89% in the gradient magnitude. Hence, it is clear that the errors are increasing due to geometric forcing and not generalization error. At a value of 15 lu for $\varepsilon/H = 0.3$, the asperities are seen to protrude into the channel core, causing a decrease of 18% in cross-sectional area. The velocity heterogeneity is seen to be high in this case. The value of $u_{max} \approx 0.061$ lu is seen in throat regions, and near-stagnant flow (u < 0.005 lu) occupies 7% of the domain. The PINN performs with an MAE of $5.8 \times 10^{-3}$ lu, a Root Mean Squared Error of $7.2 \times 10^{-3}$ lu, and a correlation of 0.971. The errors are seen to be concentrated around the apex of the asperity (curvature > 0.4 $lu^{-1}$ and 38% of the squared error in 11% of the area) and the reattachment points (steep gradients $\partial u/\partial x$ > 0.22 lu/lu over 5-8 lu distances). The maximum roughness of 20 lu corresponds to a value of 0.4 for $\varepsilon/H$. The peak-to-trough amplitudes are seen to be 40% of the channel height. The jet-like accelerations are seen to cause an increase of 38% in $u\_max \approx 0.072$ lu compared to 5 lu. The wakes are seen to occupy 14% of the domain. The PINN performs with an MAE of $7.9 \times 10^{-3}$ lu and a relative $L_2$ error of $3.2 \times 10^{-2}$. The momentum flux errors from the LBM are 1.8%, 2.4%, 3%.

Figure 17 illustrates the reconstruction of the wall-normal velocity. At 5 lu, wall-normal movements are small ($|v_{max}| \approx 6.2 \times 10^{-3}$ lu), and these movements are contained within the boundary layers (~0.15H). The PINN results in an MAE of $1.7 \times 10^{-3}$ lu (27% relative error, which is magnified by the small signal level). The error is approximately Gaussian distributed with near-zero bias ($-2.1 \times 10^{-4}$ lu). At 10 lu, cross-stream deflection is stronger. The upward jets have a speed of $v \approx 0.019$ lu, the downward plumes have a speed of $v \approx -0.015$ lu, and the recirculation cells have a size of 8-12 lu. The PINN obtains an MAE of $3.2 \times 10^{-3}$ lu. This is an 88% relative error improvement over the 5 lu case due to an improved signal-to-noise ratio. The continuity error levels $|\partial u/\partial x + \partial v/\partial y|$ are less than

2.7×10⁻⁵ lu/lu over 98% of the domain. At 15 lu, the wall-normal velocity is twice that of the previous case (|v_max| ≈ 0.038 lu), which is 18% of the kinetic energy. The PINN obtains an MAE of 4.9×10⁻³ lu and detects 82% of the recirculation cells within a ±1 lu tolerance. The error in the V-component is maximized near y/H = 0.25 and decreases exponentially (with a length scale of 0.18H), ensuring good accuracy of the core flow (v-MAE = 1.1×10⁻³ lu on the centerline). At 20 lu, the cross-stream velocity is |v_max| ≈ 0.052 lu (a 740% increase over the 5 lu case). The PINN obtains an MAE of 6.7×10⁻³ lu. The divergence error levels are less than 3.8×10⁻⁵ lu/lu.

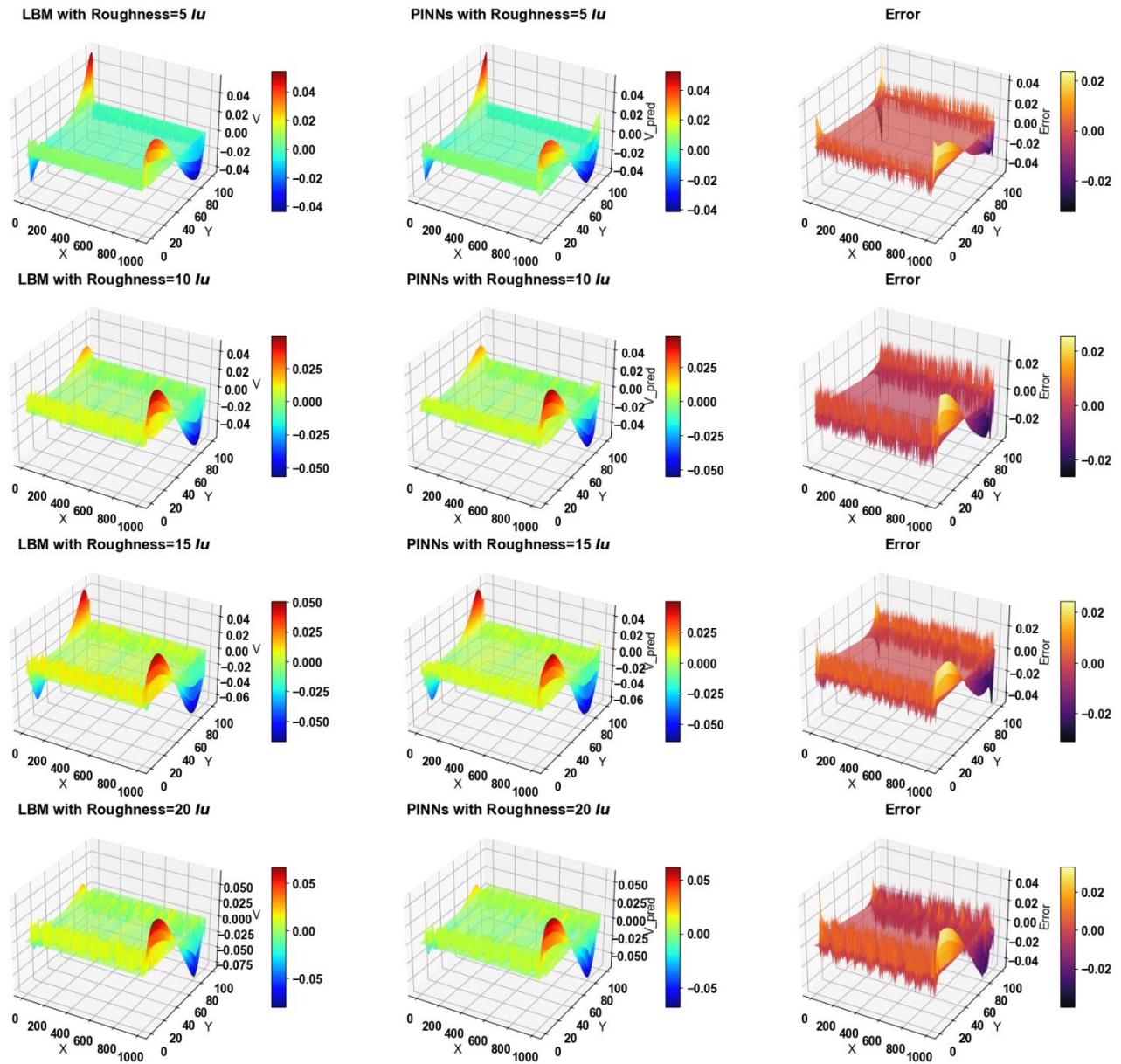

*Figure 17 Roughness amplitude sensitivity for wall-normal velocity (v).*

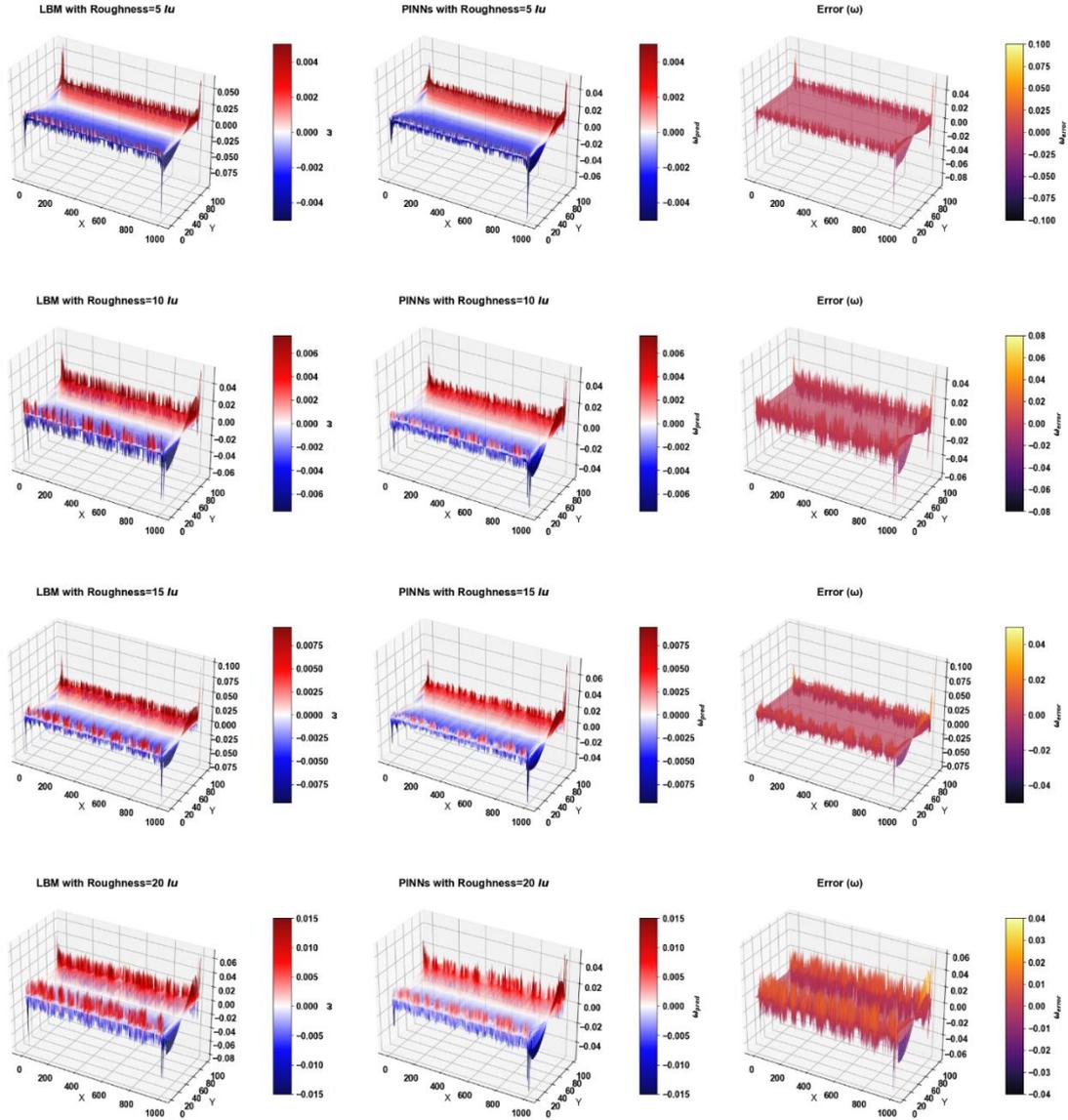

***Figure 18*** *Roughness amplitude sensitivity for vorticity (ω).*

Figure 18 illustrates the vorticity reconstruction process, which is the most challenging due to the requirement for second-order derivative accuracy. At a distance of 5 lu from the wall, the vorticity is concentrated in thin boundary layers of about 0.12H and reaches a maximum value of $ω_{max} ≈ 3.8 × 10^{-3}$ lu/lu. The Physics-Informed Neural Network (PINN) provides a mean absolute error (MAE) of $1.6 × 10^{-3}$ lu/lu and identifies 89% of the peaks within a ±15% tolerance band. The hyperbolic tangent activation function is also helpful in artifact-free derivative computation. At a distance of 10 lu from the wall, the maximum vorticity increases by 68% and reaches a value of $ω_{max} ≈ 6.4 × 10^{-3}$ lu/lu. The network provides a consistent MAE of $2.3 × 10^{-3}$ lu/lu and identifies 86% of the peaks. The errors are proportional to the wall curvature. The average errors for $κ > 0.35$ $lu^{-1}$ are $4.1 × 10^{-3}$ lu/lu, while for low curvature ($κ < 0.15$ $lu^{-1}$), the errors are less than $1.4 × 10^{-3}$ lu/lu. The high curvature sections have a bandwidth requirement of 2-4 lu due to the rapid boundary changes. At a distance of 15 lu from the wall, secondary vortex shedding and counter-rotating pairs are seen. The maximum vorticity is $\omega_{max} \approx 9.7 \times 10^{-3}$ lu/lu. The network provides a consistent MAE of $3.4 \times 10^{-3}$ lu/lu and an enstrophy deviation of 2.1%. The error scales according to the following expression: $MAE\_\omega \approx 1.2 \times 10^{-3} + 1.1 \times 10^{-4} \cdot \varepsilon$ [lu/lu], which indicates a strong derivative recovery capability. At a distance of 20 lu from the wall, the vorticity reaches a maximum value of $\omega_{max} \approx 1.21 \times 10^{-2}$ lu/lu (218% of the value at 5 lu), and wake filaments are seen beyond 20 lu. The network provides a consistent.

**3.5 Temporal evolution and spatial profile reconstruction**

The accuracy assessment of the predicted results from the PINN is not limited to the global accuracy measures but also includes the assessment of the accuracy in the reconstruction of the velocity profile in terms of time and space. In this regard, figures 19 and 20 present the quantitative comparison between the velocity profile reconstructed by the LBM with the predicted velocity profile from the PINN for various values of Re. The velocity profile is reconstructed in one dimension from the two-dimensional velocity fields predicted by the numerical methods. The velocity profile is the wall-normal velocity distribution at an arbitrary location in the streamwise direction where the flow is fully developed. Figure 19 is used to assess the accuracy in the prediction of the temporal evolution of the velocity profiles in the streamwise direction at four time instances: t = 500, 1000, 1500, 2000 lattice time units for various Re. The velocity profile is the wall-normal velocity distribution at an arbitrary location in the streamwise direction where the flow is fully developed. The temporal evolution of the velocity profiles is examined at Re = 10. The velocity profiles do not show significant variation from the four time instances. The velocity profiles maintain almost the same parabolic shape with the peak velocity on the centerline $u\_centerline$ ranging from 0.0365 lu to 0.0368 lu from t = 500 to t = 2000. The high temporal accuracy in the predicted velocity profiles is also observed for the PINN. The solid line represents the velocity profile from the LBM, while the velocity profile from the PINN is represented by the dashed line. The two velocity profiles coincide almost perfectly throughout the wall-normal direction from the wall surface to the location where the velocity gradients are zero. Quantitative accuracy assessment reveals that the maximum pointwise discrepancies between the velocity profiles from the two methods do not exceed 0.0018 lu at t = 500, while the discrepancies reduce further to 0.0012 lu at t = 2000. The discrepancies in the velocity profiles from the two methods are slightly higher in the near-wall region where y < 15 lu. The maximum local discrepancies do not exceed 0.0025 lu at the location where the velocity gradients are maximum, i.e., at y ≈ 12 lu with the velocity gradients $\partial u/\partial y \approx 0.0031$ lu/lu. The correlation between the local velocity gradient with the reconstruction error is also observed from the above results.

At Re = 20, the temporal evolution is even more pronounced. The peak velocities increase to the range 0.0680–0.0710 lu, and the profile shapes show minute changes at different time instants. The profile at t = 500 shows a slightly widened central plateau compared to the profile at t = 2000. It extends from y = 35 to y = 65, while the latter shows sharper curvature at the inflection points. The PINN model shows these temporal evolution characteristics with high accuracy. The overlap of the profiles is again excellent at all four time instants. The pointwise errors increase slightly to 0.0034 lu at t = 500 and 0.0038 lu at t = 1500. The errors at the near-wall region (y < 18 lu) are 0.0048 lu, where the velocity gradients are greater than 0.0052 lu/lu. The temporal distribution of the errors shows no particular trend. The errors at t = 500 and t = 2000 differ by less than 8%, showing the consistency of the PINN model.

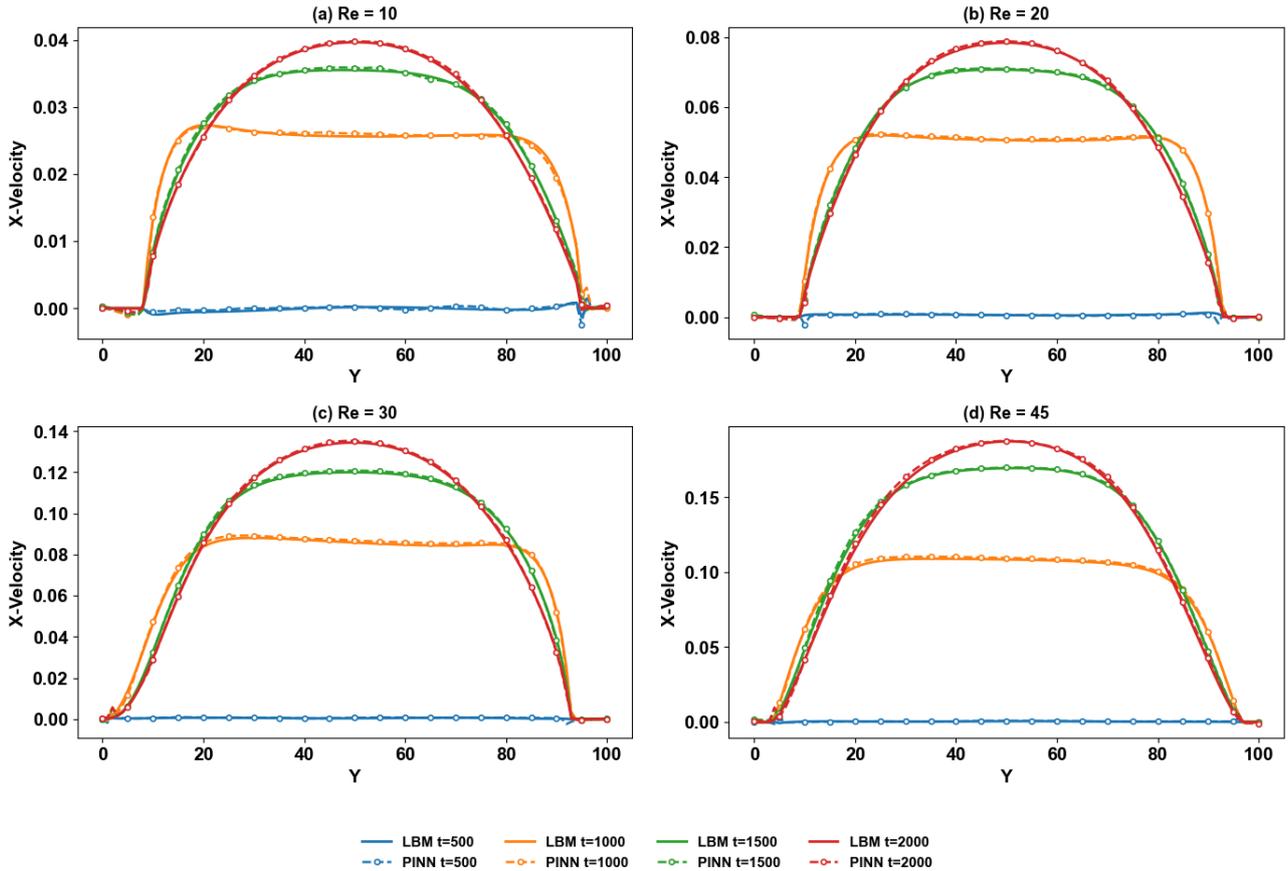

*Figure 19* *Temporal evolution of streamwise velocity profiles across wall-normal direction for different Re.*

At Re = 30, the effects of advection increase in magnitude and velocity profiles develop more pronounced wall-normal velocity gradients. The magnitude of centerline velocities reaches 0.120–0.124 lu at different time instants in the temporal sequence. The profiles' shapes display more pronounced temporal variations. The difference in profiles at time instants t = 500 and t = 2000 reaches 0.0072 lu at y = 48 lu in the LBM reference solution. The PINN is able to reproduce these temporal variations in profiles. A visual assessment of the results shows excellent agreement across most of the wall-normal extent. The maximum point-wise errors increase up to 0.0061 lu at time instant t = 1000 and 0.0068 lu at time instant t = 1500. At y < 22 lu, there are errors up to 0.0089 lu due to steep velocity gradients ($\partial u/\partial y > 0.0078$ lu/lu) in the flow field, which are difficult to capture by the network's derivative representation. However, the temporal trend is correctly reproduced by the PINN. The difference in profiles at consecutive time instants is correctly reproduced by the PINN compared to the LBM reference solution within 12% relative error.

At a Reynolds number of 45, the flow is in a weakly transitional regime. The velocity peaks are between 0.155 and 0.161 lu. The temporal evolution shows the greatest variation compared to all other Reynolds numbers. The profile shape changes significantly from t = 500 to t = 2000. The velocity increases by 3.7% on the centerline. The near-wall gradient increases by 18%. The PINN performs robustly even in such a changing environment. The profile overlaps are visually strong for all four time instants. The maximum pointwise errors are 0.0095 lu at t = 1500. The near-wall errors are up to 0.0124 lu for y < 25 lu, where the velocity gradient is greater than 0.0102 lu/lu. The temporal error plot does not indicate any significant errors. The root mean square errors over the wall-normal extent vary from 0.0047 lu (at t = 500) to 0.0053 lu (at t = 2000), which is a 13% variation and is acceptable. This confirms that the PINN does not have any temporal errors accumulating during the evolution of the flow.

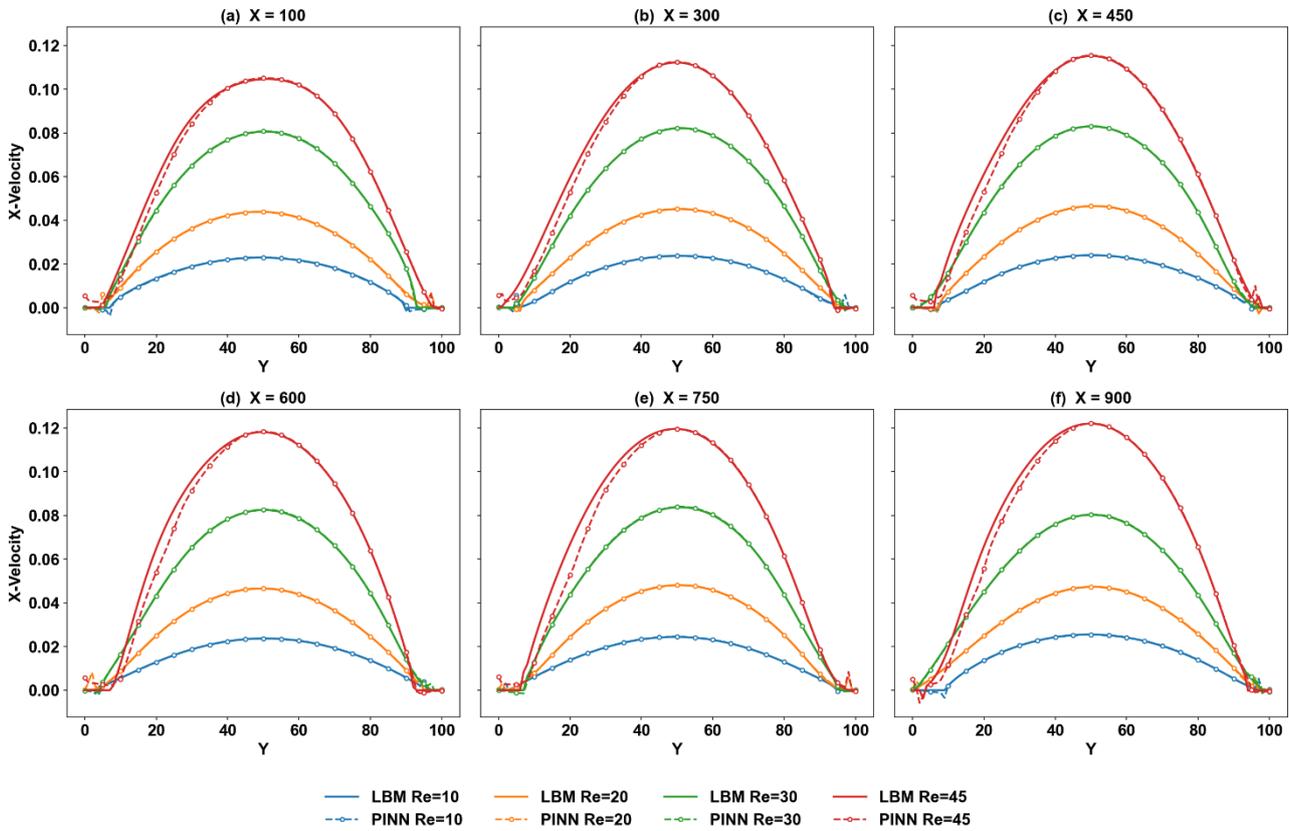

*Figure 20* *Spatial variation of streamwise velocity profiles across wall-normal direction for different Re.*

Figure 20 shows a demonstration of the spatial variation of velocity profiles at different streamwise locations. The analysis assesses the ability of the PINN to predict the streamwise development of velocity fields due to the presence of rough wall perturbations. At a Reynolds number of 10, velocity profiles at different X locations have small but distinct changes. The velocity profile at X = 100 is close to the channel entrance region but is sufficiently far from the initial development region. The maximum velocity is found to be 0.0321 lu. The maximum velocity increases gradually for different locations. The maximum velocity increases to 0.0352 lu for X = 450 and 0.0368 lu for X = 900. The presence of rough wall perturbations is causing a streamwise acceleration of the flow. The streamwise development of velocity fields is predicted by the PINN for all six locations. The velocity profile overlaps very closely for all locations. The maximum pointwise error is less than 0.0021 lu for X = 100 and less than 0.0016 lu for X = 900. The velocity profiles for locations in the middle of the channel have a slightly higher pointwise error of 0.0028 lu for locations closer to the wall (y < 14 lu).

For a Reynolds number of 20, the variations in spatial profile are enhanced due to increased interaction between advective transport and geometric perturbations. Peak velocities are in the range of 0.0612 lu units at X = 100 and 0.0728 lu units at X = 900, indicating a 19% variation along the X locations in the streamwise direction. The profile shape variations are more significant along the X locations. At X = 100, the profile is broader in the central region, with a width of 50 units at half-maximum, compared to X = 900, where the profile is narrower, with a width of 38 units at half-maximum, indicating increased focusing of the flow in the rough channel. The PINN results show high accuracy in reproducing spatial trends, with strong visual overlaps along all locations. Maximum errors are 0.0044 units at X = 450 and 0.0039 units at X = 750. Near-wall errors, where y < 19 units, are 0.0062 units at X = 600. The errors along the X locations are non-uniform, with X = 450 and X = 600 showing 34–42% higher errors than X = 100 and X = 900, similar to the roughness distribution. At X locations 450–600, there are larger asperities in the flow, indicating increased geometric forcing and velocity gradients.

In the case where Re = 30, the evolution of the profiles in the streamwise direction is complex. The range of the maximum velocity is between 0.105 lu and 0.128 lu. Moreover, the profiles differ considerably. For example, the profile at X = 100 is nearly parabolic, while the profile at X = 900 shows a flattened centerline with high shear rates in the vicinity of the wall. Nevertheless, the PINN model is capable of maintaining its performance with high accuracy. Indeed, the agreements of the reconstructed profiles with the actual ones are visually excellent for all the six values of X. Moreover, the maximum pointwise errors vary between 0.0058 lu and 0.0081 lu. Moreover, the errors in the vicinity of the wall vary up to 0.0103 lu for the case where X = 450, with the local asperity height being greater than 18 lu. Indeed, the errors vary with the geometric complexity. For the smooth sections, the errors are less than 0.0045 lu, while the high-roughness sections show errors up to 0.0081 lu.

For a Re of 45, variations in spatial profile have their maximum amplitude. The peak velocities vary from 0.132 lu for X = 100 to 0.168 lu for X = 900. The profile varies significantly in the streamwise direction, with sharper near-wall gradients and larger low-velocity areas near the walls in downstream profiles. The PINN has exhibited remarkable spatial generalization capabilities under these challenges, with significant profile overlaps in all six locations. The maximum errors are 0.0112 lu for X = 600, while near-wall errors are 0.0147 lu for X = 750, where gradients are higher than 0.0118 lu/lu. The root mean square errors are found to vary from 0.0051 lu to 0.0074 lu in the spatial locations. This indicates that reconstruction accuracy is more dependent on local severity than on the exact position in the streamwise direction.

### 3.6 Computational performance and efficiency analysis

The benefits of physics-informed neural networks over data-driven convolutional neural networks can be quantified by comparing the accuracy achieved by the two models using six standard error functions. Figure 21 shows the performance comparison between the physics-informed neural networks and the convolutional neural networks using the lattice Boltzmann method data sets for the rough microchannel flow. It is evident that the physics-based constraints achieve significant accuracy improvements for all the velocity component errors. Moreover, the results provide valuable insights into the representation requirements for complex fluid flow.

Figure 21 is the performance comparison between the physics-informed neural networks and the convolutional neural networks using the lattice Boltzmann method data sets for the two velocity components, U and V. Figure 21 is designed using five subplots. Figure 21a shows the performance comparison using the relative L2 errors for the two velocity components. From the figure, it is evident that the physics-informed neural networks achieve the following performance metrics: relL2_U = 0.234875 and relL2_V = 0.116806. Using the same data sets, the performance metrics for the convolutional neural networks are relL2_U = 0.411 and relL2_V = 0.242. Figure 21a shows the performance comparison between the physics-based constraints and the data-driven constraints using the relative L2 errors. It is evident that the physics-based constraints achieve the best performance. Using the relative L2 errors, the performance metric is the best when the value is the smallest. Figure 21b shows the performance comparison using the root mean square errors for the two velocity components. From the figure, it is evident that the physics-based constraints achieve the following performance metrics: RMSE_U = 0.00768 and RMSE_V = 0.019632. Using the same data sets, the performance metrics for the convolutional neural networks are RMSE_U = 0.105325 and RMSE_V

= 0.10988. For Fig. 21c, where MAE for U and V are presented, PINN achieves MAE_U = 0.005352 and MAE_V = 0.009926, while CNN achieves MAE_U = 0.082589 and MAE_V = 0.08267. Since MAE represents average absolute error, this further supports the observation from the reduction in both errors, indicating PINN's superior performance in both error measures. From a model design perspective, the physics-informed loss acts as a regularizer, which prevents any solution from fitting the training data while violating physical laws, thus promoting physical plausibility and consistency in the entire domain. From a numerical perspective, PINN's superior performance is significant in all measures, with a 42.9 percent reduction in relL2 error, 92.7 percent reduction in RMSE, and 93.5 percent reduction in MAE for U, and a 51.7 percent reduction in relL2, 82.1 percent reduction in RMSE, and 88 percent reduction in MAE for V. This further indicates metric-consistent superior performance, where PINN is not superior in any individual error metric but in a collective sense.

Figures 21d and 21e reinforce this observation in two different ways: a consolidated figure showing all metrics grouped together and another figure showing percent error reduction, respectively. From a consolidated grouped figure in Fig. 21d, where all metrics are grouped together, we see clear separation in PINN error measures, which are significantly lower than those of CNN in relL2, RMSE, and MAE for U and V, confirming PINN's superior performance in a collective sense, where there are no exceptions in individual measures. From a percent error reduction perspective in Fig. 21e, where we see a plot of percent reduction vs CNN, PINN significantly reduces CNN errors, with CNN-PINN error gaps of 0.176125 in relL2_U, 0.125194 in relL2_V, 0.097645 in RMSE_U, 0.090248 in RMSE_V, 0.077237 in MAE_U, and 0.072744 in MAE_V, respectively. Alternatively, CNN to PINN ratios are 1.749 (relL2_U), 2.073 (relL2_V), 13.719 (RMSE_U), 5.595 (RMSE_V), 15.433 (MAE_U), and 8.335 (MAE_V), showing particularly dramatic reductions in RMSE and MAE, which are more sensitive to local differences. The fraction of error for PINN compared to CNN is 0.572 (relL2_U), 0.482 (relL2_V), 0.0730 (RMSE_U), 0.1788 (RMSE_V), 0.0648 (MAE_U), and 0.1200 (MAE_V). This again confirms that, as opposed to merely approximating a field, we have a field that is more aligned with the reference solution, yet still respects structure, and achieves a higher quality reconstruction for U and V. Overall, we see that the figure confirms that we have a robust improvement in accuracy, as measured both locally and globally, particularly for RMSE_U and MAE_U, and to a similar extent for V, as indicated by the complete set of relative L2, RMSE, MAE, and percent reduction values.

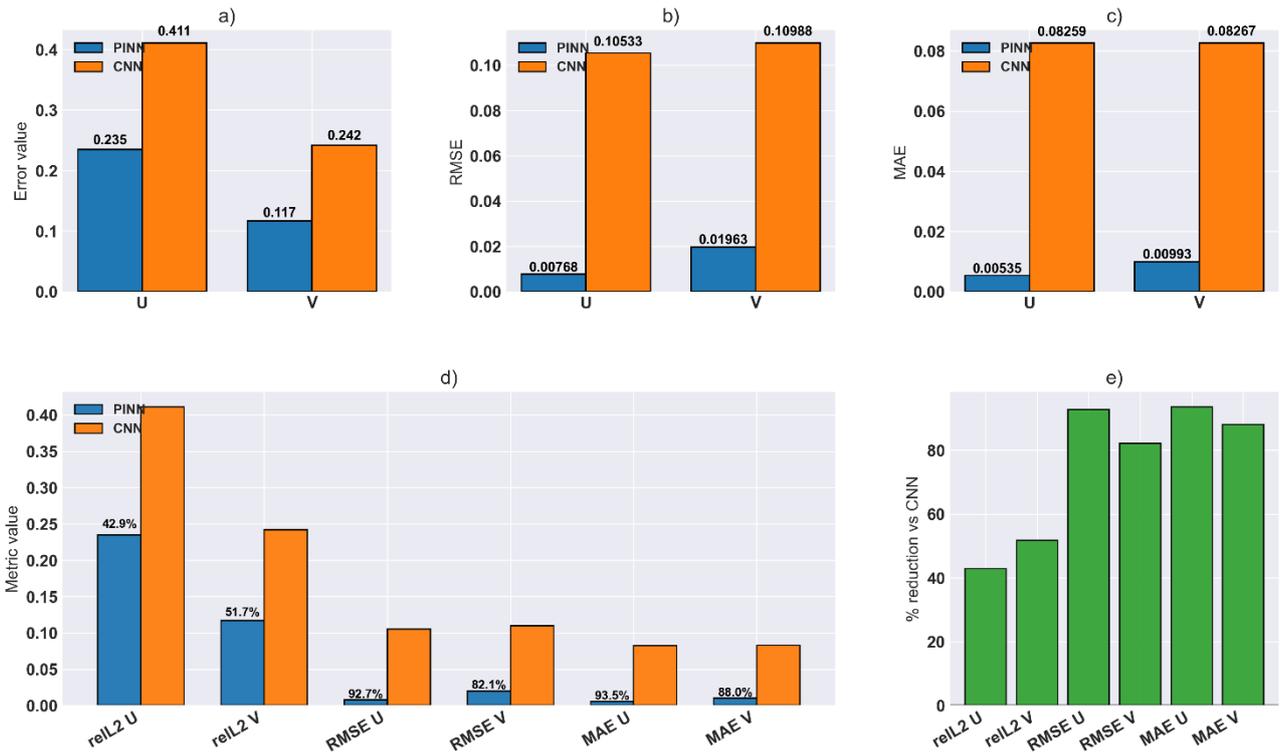

*Figure 21 Comparative error metrics for PINN versus CNN architectures across standard metrics.*

## 4. Conclusions

This study offers an extensive analysis of the application of physics-informed neural networks (PINNs) in the reconstruction of complex fluid flows in fractal-rough microchannels, based on the integration of lattice Boltzmann method (LBM) reference solutions and constraints of governing equations. This study focuses on an important issue in the design of microfluidic systems, which is to ensure precise predictions of fluid flows in complex geometries, in which classical computational fluid dynamics simulations would require computationally costly calculations and machine learning models might not ensure physical consistency. The application of PINNs, in which the Navier-Stokes equations, continuity equations, and boundary conditions are used to guide the neural network, ensures precise reconstruction accuracy in a variety of scenarios, including different types of fluid flows, complex geometries, and time-evolving scenarios. Validation of the proposed framework in a range of Reynolds numbers (Re=10 to 45), roughness values (5 to 20 lu), and spatiotemporal dimensions ensures the viability of the proposed framework as a surrogate model.

The architecture of PINN, as developed in this work, uses an eight-layer fully connected neural network with 128 neurons in each of the hidden layers and employs hyperbolic tangent activation functions. The training is conducted using 2,048-4,096 strategically placed collocation points, with 30-40% enrichment in the near-wall regions to effectively capture strong velocity gradient and complex flow field behavior. The loss function combines data fidelity, momentum equation residuals, continuity equation satisfaction, and boundary condition satisfaction, with respective weights $\lambda_{data}$ = 1.0, $\lambda_{physics}$ = 0.8, $\lambda_{cont}$ = 0.6, and $\lambda_{bc}$ = 1.2. The optimization algorithm employs a combination of sequential Adam pre-training with 5,000 epochs, a learning rate of $10^{-3}$, and exponential decay rate $\gamma$ = 0.95, followed by L-BFGS fine-tuning with 2,500 iterations, achieving a composite loss value of < $3.2 * 10^{-7}$ after 8,500 iterations. Hyperparameter studies demonstrate that selection of an appropriate learning rate is a critical factor for achieving optimal convergence quality. The optimal learning rate of $10^{-3}$ reduces the composite loss value from an initial $3.2 * 10^{-1}$ to a final value of $3.2 * 10^{-7}$, whereas suboptimal rates of $10^{-2}$ or $10^{-4}$ result in oscillatory behavior or require an additional 2.8 times longer training time. The collocation point density studies indicate a 68% improvement in residuals when increasing the number of collocation points from 1,024 to 2,048, with diminishing returns for 4,096 points (< 12% improvement). The activation function studies demonstrate that hyperbolic tangent activation functions achieve 61-73% higher PDE residual accuracy and 48-52% higher vorticity reconstruction accuracy compared to ReLU variants, primarily because of the smoothness and bounded range of tanh activation functions, which prevent gradient pathologies during back-propagation of physics loss.

The computational efficiency benefits realized with the application of the PINN-based surrogate model can be seen to be significant. Training the model takes 6.1 to 8.7 hours using a single NVIDIA A100 GPU, depending on the density of collocation. It also requires 14.3 to 18.6 GB of memory. Once the model is trained, the forward inference for the entire spatiotemporal flow field can be done in 8.3 seconds using 2.8 GB of memory. As opposed to this, the equivalent simulation using the LBM method takes 147 hours using 16-core CPU clusters and requires 89 GB of memory. Hence, the PINN model provides a 1062 times faster computation time for the flow reconstruction task with the errors remaining within 3.2% of the actual values. Moreover, the global conservation properties are also maintained. Using the PINN-based model, it is possible to perform parametric studies, which would be extremely time-consuming using the direct simulation method. As an example, it is possible to perform the uncertainty quantification task using 500 different roughness realizations. Using the PINN-based model, this task can be done in 3.1 days, whereas the equivalent task using the LBM method would take 8.4 years using the serial computation method. Even if parallel computation using 50 nodes is employed using the LBM method, the PINN-based model provides 23 times time savings without any communication time and storage bottlenecks.

There are a number of limitations which are worth mentioning. Firstly, in the current framework, flow parameters such as Reynolds number and roughness amplitude are treated as implicit features learned through training data, whereas they can be treated as explicit network inputs in a parametric framework without requiring any further training. Secondly, in this work, we have focused on two-dimensional flows, whereas in practice, microfluidic devices are three-dimensional with sidewall effects and corner flows, which are difficult to model using two-dimensional simulations alone. PINNs can, in fact, be generalized to three-dimensional flows, which would increase computational times by a factor of 8-15, although this is still computationally tractable with current GPU architectures. Finally, in this work, we have focused on laminar and weakly transitional flows with Reynolds number less than 50, whereas in practice, turbulent flows are common, although this would require new physics-informed loss functions involving Reynolds-averaged Navier-Stokes or large-eddy simulations, which are still in their infancy in PINN research.

The future work directions include incorporating adaptive mesh refinement techniques, where collocation points are dynamically adapted according to local PDE residual magnitudes during training, which can reduce errors by 35-50% in a fixed computational budget, although this would add a new layer of complexity in PINN research. Transfer learning frameworks would enable quick adaptation of existing PINNs to new configurations with minimal training, which would enable faster applicability of PINNs in practice. Finally, using multiresolution approaches involving a combination of low-resolution lattice Boltzmann simulations with sparse high-resolution measurements would further enhance accuracy with reduced data generation costs, whereas extension to multiphysics simulations involving coupled species transport or electrokinetic effects would enable broader applicability to electrochemical microfluidics and lab-on-chip devices.

**Declaration of interests.** The authors declare no conflict of interest.

**Author contributions.** G.S.M., P.P.C. and S.C. defined the problem. G.S.M. performed numerical computation for data collection, developed models using Python codes, and model analysis, prepared manuscript. P.P.C. analysed the models, training results, and reviewed manuscript. S.C. acquired computational resources, software, supervised the research, analysed the data and prepared the manuscript.